\documentclass[10pt,reqno]{amsart}
\usepackage{amssymb}
\usepackage{amsfonts}
\usepackage{amsthm}
\usepackage{amsmath}

\newcommand{\Reals}{{\mathbb{R}}}
\newcommand{\Cmplx}{{\mathbb{C}}}
\newcommand{\Ints}{{\mathbb{Z}}}
\newcommand{\Disk}{{\mathbb{D}}}

\newcommand{\C}{{\mathcal{C}}}

\let\det=\undefined\DeclareMathOperator*{\det}{det}
\let\Re=\undefined\DeclareMathOperator*{\Re}{Re}
\let\Im=\undefined\DeclareMathOperator*{\Im}{Im}

\DeclareMathOperator*{\tr}{tr}
\DeclareMathOperator*{\diag}{diag}
\DeclareMathOperator*{\supp}{supp}
\DeclareMathOperator*{\Id}{Id}
\DeclareMathOperator{\Exp}{{{\mathbb{E}}}}
\DeclareMathOperator{\Prob}{{{\mathbb{P}}}}

\def\e{\mathfrak{e}}
\def\vp{{\vec{\smash{\psi}}}}
\def\vpn{\vp^{{}_{(\!n\!)}}}

\newtheorem{theorem}{Theorem}[section]
\newtheorem{prop}[theorem]{Proposition}
\newtheorem{lemma}[theorem]{Lemma}
\newtheorem{coro}[theorem]{Corollary}
\theoremstyle{definition}
\newtheorem{definition}[theorem]{Definition}
\theoremstyle{remark}

\newtheorem*{warn}{Warning}

\begin{document}

\title[Eigenvalue Statistics for CMV matrices]{Eigenvalue Statistics for CMV Matrices:\\%
From Poisson to Clock via C$\beta$E.}
\author[R.~Killip and M. Stoiciu]{Rowan Killip and Mihai Stoiciu}
\address{Rowan Killip\\
         UCLA Mathematics Department\\
         Box 951555\\
         Los Angeles, CA 90095}
\email{killip@math.ucla.edu}
\address{Mihai Stoiciu\\
    Department of Mathematics and Statistics\\
    Williams College\\
    Williamstown, MA 01267}
\email{mstoiciu@williams.edu}

\date{\today}

\begin{abstract}
We study CMV matrices (a discrete one-dimensional Dirac-type operator) with random decaying coefficients.
Under mild assumptions we identify the local eigenvalue statistics in the natural scaling limit.
For rapidly decreasing coefficients, the eigenvalues have rigid spacing (like the numerals on a clock);
in the case of slow decrease, the eigenvalues are distributed according to a Poisson process.  For a certain
critical rate of decay we obtain the circular beta ensembles of random matrix theory.  The temperature $\beta^{-1}$
appears as the square of the coupling constant.
\end{abstract}

\maketitle

\section{Introduction}

The bulk spectral properties of Schr\"odinger operators and Jacobi matrices with random decaying
potentials are fairly well understood; see \cite{DELYON,DELYON-S-S,KLS,KOTANI-USHIROYA,SIMON-SJM}.
In particular, it has been discovered that there is a change in spectral type at certain critical
rate of decay.  It is natural to investigate the finer spectral properties; for example, the
distribution of eigenvalue spacings.  In order to speak of eigenvalue spacings, one first restricts
the operator to a finite volume and then studies the behaviour of the rescaled spacings in the
infinite volume limit.  Such an investigation is the goal of this paper; however we treat neither
Schr\"odinger nor Jacobi operators, but rather CMV matrices:

\begin{definition}\label{D:CMV}
Given an infinite sequence of coefficients $\alpha_0,\alpha_1,\ldots$ in $\Disk$ (the open
unit disk in $\Cmplx$), we define $2\times 2$ matrices
$$
\Xi_k = \begin{bmatrix} \bar\alpha_k & \rho_k \\ \rho_k & -\alpha_k
\end{bmatrix},
$$
where $\rho_k=\sqrt{1-|\alpha_k|^2}$. From these, we form block-diagonal matrices
$$
\mathcal{L}=\diag\bigl(\Xi_0   ,\Xi_2,\Xi_4,\ldots\bigr)
\quad\text{and}\quad
\mathcal{M}=\diag\bigl(\Xi_{-1},\Xi_1,\Xi_3,\ldots\bigr),
$$
where $\Xi_{-1}=[1]$, the $1\times 1$ identity matrix.
The \emph{CMV matrix} associated to the coefficients $\alpha_0,\alpha_1,\ldots$ is
$\C=\C(\alpha_0,\alpha_1,\ldots)=\mathcal{LM}$.
\end{definition}

The name CMV is taken from from the initials of the authors of \cite{CMV}, following recent custom; however
as discussed in \cite{SimonCMV5,Watkins}, they appeared earlier in several places.  A thorough discussion of
these operators can be found in \cite{SimonOPUC1,SimonOPUC2}.

CMV matrices are unitary and may naturally be considered a discrete form of one-dimensional Dirac operator.  There are
several reasons for our choosing to consider these operators: they are bounded, discrete, and most importantly,
possess a rotational symmetry that we will discuss in due course.

The restriction of a CMV matrix to a finite interval is not simply the top left corner of the original matrix---this
would result in a non-unitary operator---one needs to add a boundary condition.  If one replaces $\alpha_{n-1}$ by
$e^{i\eta}\in S^1$ (the unit circle in $\Cmplx$), then $\C$ decomposes into a direct sum; specifically,
the top left $n\times n$ block decouples from the remainder of the matrix.  We define the $n\times n$ truncation
of $\C$ with $\eta$ boundary condition in this way and write
$$
\C^{(n)} = \C(\alpha_0,\ldots,\alpha_{n-2},e^{i\eta}).
$$

We wish to consider the model where $e^{i\eta}$ and each $\alpha_k$ is chosen independently
(though not identically distributed) with a rotationally invariant distribution, that is,
$$
\Exp\Bigl\{ g(e^{i\eta})\prod f_k(e^{i\theta_k}\alpha_k) \Bigr\}
= \int_0^{2\pi} g(e^{i\eta}) \,\tfrac{d\eta}{2\pi} \ \prod  \Exp\{ f_k(\alpha_k)\}
$$
for any finite collection of $f_k:\Disk\to\Cmplx$ and $\theta_k\in\Reals$.  The reason for considering
only rotationally invariant distributions is that in this case the spectral properties are homogeneous under
rotations of the circle.  For example, for any $\theta\in[0,2\pi)$,
\begin{equation}\label{RotInvar}
e^{-i\theta}\C(\alpha_0,\alpha_1,\ldots,\alpha_{n-2},e^{i\eta}) \cong
\C(e^{i\theta}\alpha_0,e^{2i\theta}\alpha_1,\ldots,e^{i(n-1)\theta}\alpha_{n-2},e^{i\eta+in\theta}),
\end{equation}
which is to say that the two matrices have the same eigenvalues that is, they are unitarily equivalent
(cf. \cite[p. 81]{SimonOPUC1}).

For power-decaying $\alpha_k$, the bulk spectral properties for this model were worked out in
\cite[\S 12.7]{SimonOPUC2}, following the lines of earlier arguments from the Jacobi and
Schr\"odinger cases. Let us now describe the finer behaviour we wish to investigate.

Associated to the random $n\times n$ matrix $\C^{(n)}$ are $n$ random eigenvalues,
which we denote $e^{i\theta_1},\ldots,e^{i\theta_n}$, with $\theta_j\in(-\pi,\pi]$.
By rotation invariance, \eqref{RotInvar},
the average number of eigenvalues with $\theta_j\in[a,b)$ is $\tfrac{b-a}{2\pi}n$.  For this reason it is
natural to rescale the eigenvalues by a factor of $n$.  Our goal is to show that under natural assumptions,
the \emph{eigenvalue process}, that is, the random set of points $\{n\theta_1,\ldots,n\theta_n\}\subset\Reals$,
converge (in an appropriate sense) to a point process and to understand the limiting process as best we can.

We will prove the convergence of the expected values of certain random variables that are rich enough to
characterize the point process:

\begin{definition}\label{D:TestRV}
Given a non-negative $f\in C^\infty_c(\Reals)$, we define a random variable on a point process by
$$
 X_f = \exp\Bigl( - \sum f(x_j) \Bigr)
$$
where $x_j$ are the points of the process.  $\Exp\{X_f\}$ are known as the Laplace functionals of the
process.
\end{definition}

One may also characterize a point process in terms of the probability of certain events:

\begin{definition}\label{D:TestEvent}
Given $p$ disjoint intervals $[a_j,b_j)$, and $p$ non-negative integers $k_1,\ldots,k_p$, we define the event
$$
\Omega\bigl([a_1,b_1),\ldots,[a_p,b_p),k_1,\ldots,k_p\bigr)
$$
as the occurrence of exactly $k_j$ points in the interval $[a_j,b_j)$ for each $1\leq j\leq p$.
\end{definition}

Convergence of the probabilities of these events is fractionally stronger than that of the random variables $X_f$;
indeed they are equivalent notions if no location is occupied by the process with positive probability
(cf. \cite[\S3.5]{Resnick}). We will use the former test for convergence as it is the most convenient for us.

Three processes will arise as limits of the eigenvalue process; we will now define these.

\begin{definition}\label{D:Poisson}
The Poisson process with intensity $d\nu$ is characterized by
$$
\Exp\bigl\{ X_f \bigr\} = \exp\biggl\{ \int \bigl[e^{-f(x)}-1\bigr]\,d\nu(x) \biggr\}
$$
or more simply by
$$
\Prob\bigl\{ \Omega\bigl([a_1,b_1),\ldots,[a_p,b_p),k_1,\ldots,k_p\bigr) \bigr\}
    = \prod_{j=1}^p \frac{\lambda_j^{k_j} e^{-\lambda_j}}{k_j!}
$$
where $\lambda_j=\nu([a_j,b_j))$.  (This says that the number of points in disjoint intervals
are independent Poisson random variables.)
\end{definition}

\begin{definition}\label{D:Clock}
The clock (or picket fence) process with spacing $h$ is characterized by
$$
\Exp\bigl\{ X_f \bigr\} = \frac{1}{h} \int_0^h \exp\Bigl\{{\textstyle -\sum_{k\in\Ints} f(hk+x) }\Bigr\} \,dx.
$$
That is, points occur a random translate of $h\Ints$.
\end{definition}

The third limiting process we need to describe is considerably more complex.  We begin with the
C$\beta$E${}_n$, that is, the circular $\beta$ ensemble with $n$ points.  This is a point
process on the unit circle.  The total number of points is $n$ (non-random) and they are distributed so that
\begin{align}\label{CGbeta}
\Exp_n^\beta (f) &=
\frac{1}{Z_{n,\beta}} \int_{-\pi}^\pi\!\! \cdots  \int_{-\pi}^\pi f(e^{i\theta_1},\ldots,e^{i\theta_n})
    \bigl|\Delta(e^{i\theta_1},\ldots,e^{i\theta_n})\bigr|^\beta \, \frac{d\theta_1}{2\pi} \cdots \frac{d\theta_n}{2\pi}
\end{align}
for any symmetric function $f$. The partition function is given by
\begin{equation}\label{CGpart}
  Z_{n,\beta} = \frac{\Gamma(\tfrac12\beta n + 1)}{\bigl[\Gamma(\tfrac12\beta + 1)\bigr]^n};
\end{equation}
see \cite{Good,Gunson,Wilson}.  This is a standard family of ensembles in random matrix theory and was
introduced by Dyson, \cite{Dyson}.  This ensemble also arises as the Gibb's measure for $n$ identical charged
particles confined to lie on the circle and interacting via the $2$-dimensional Coulomb law.  For this reason,
it is also known as the $\log$-gas.

Just as we discussed earlier for eigenvalues of random CMV matrices, one may take the points of the C$\beta$E${}_n$
and by unwrapping and rescaling obtain a sequence of point processes on the line.  It is not obvious that this
sequence of point processes converges.  One off-shoot of the work presented here is that they do (indeed,
we will obtain a new representation of the limiting process).

\begin{definition}\label{D:CBE}
The C$\beta$E process is characterized by
$$
\Exp\bigl\{ X_f \bigr\} = \lim_{n\to\infty} \Exp^\beta_n\Bigl\{
    \exp\bigl\{ - {\textstyle \sum_{j} f(n\theta_j)}\bigr\} \Bigr\}.
$$
with the notation of \eqref{CGbeta}. (This limit does indeed exist.)
\end{definition}

It is not difficult to show that $\beta\downarrow 0$ limit of C$\beta$E${}_n$ converges to a Poisson
process as $n\to\infty$ and that the $\beta\uparrow \infty$ limit converges a clock process.

For $\beta\in\{1,2,4\}$, the C$\beta$E process is very well understood; indeed there are explicit formulae for
the correlation functions and for the inter-particle spacing distributions.  Discussion of this work can scarcely
be contained in a single monograph and we have no intention of trying to review it here; the unacquainted should
probably begin with \cite{Mehta} or \cite{ForresterBook}.

The content of this paper can be best summarized as

\begin{theorem}\label{T:main}
(i) If\/ $\Exp \left\{ |\alpha_k|^2 \right\}=o(k^{-1})$, then the eigenvalue process converges to the
clock process with spacing $2\pi$ as $n\to\infty$.\\
(ii) If $\Exp\left\{ |\alpha_k|^2 \right\} - \tfrac{2}{\beta(k+1)} =o(k^{-1})$
and $\Exp\left\{ \log^{2}[1-|\alpha_k|^2] \right\} =o( k^{-1} )$, then the eigenvalue process converges
to the C$\beta\!$E process as $n\to\infty$. \\
(iii) If $\Exp \left\{ |\alpha_k|^2 \right\} \geq (k+1)^{\epsilon-1}$ for some $\epsilon >0$ and
$\Exp \left\{ (1-|\alpha_k|^2)^{-s} \right\} = O(1)$ for some $s>0$,
then the eigenvalue process converges to the Poisson process with intensity $\tfrac{1}{2\pi}dx$ as $n\to\infty$.
\end{theorem}

This theorem demonstrates that (at least for this model) the transition from Poisson to Clock statistics
is actually continuous.  This moots the question of the eigenvalue statistics in a (microscopic)
neighbourhood of the (conjectured) metal/insulator transition in the multi-dimensional Anderson model.
One may also ponder the eigenvalue statistics for less random models, for example, the Almost-Mathieu
operator, which is also known to have such a transition. Unfortunately, we have nothing to say on these
interesting topics.

As $\log[1-r^2]=r^2+O(r^4)$ as $r\downarrow0$, it is natural to consider replacing the second hypothesis
in part (ii) by $\Exp\left\{ |\alpha_k|^4 \right\}=o(k^{-1})$ or perhaps discarding it altogether.  The
first option is conceivable, but would increase the complexity of the proof more than it warrants.
This second option is certainly not possible.  An $\alpha_k$ lying very close to the unit circle effectively
of decouples the CMV matrix into a direct sum.  For example, if one considers
$$
|\alpha_k| = 1 - e^{-k!} \quad\text{with probability $\sim \tfrac{2}{\beta k}$},
$$
and zero otherwise, then the eigenvalue process becomes a superposition of clock processes with random spacings.
This has too little eigenvalue repulsion to be consistent with C$\beta$E, at least for $\beta$ large.  Similarly,
the second hypothesis in part (iii) is surely an artifact of our proof.

The proof of part (i) is the most elementary of the three and is contained in Section~\ref{S:Fast} using
results from Section~\ref{S:Basic}.  A discussion of non-random results of this type
(with stronger decay assumptions) can be found in \cite[\S 4]{SimonFineI}; their proofs occupy the subsequent
sections of that paper.

Part (iii) of Theorem~\ref{T:main} is proved in Section~\ref{S:Slow} using methods from the Ph.D. thesis of
the second author, \cite{StoiciuPhD,MSjat}.  This in turn is derived from work of Minami, \cite{Minami}, for
the Anderson model.

Part (ii) is proved by combining two ingredients.  The first is an exactly soluble example of our model that
was discovered in \cite{KN}.  The second is an invariance principle that shows that the limiting eigenvalue
statistics do not depend on the specifics of the probability distribution of the Verblunsky parameters.

Several results of this genre have been proved, though only for systems which lead to $\beta\in\{1,2,4\}$.  Such results
are reviewed in \cite{DeiftUni}.  To our eyes, the most similar results of this type can be found in
\cite{Johansson,Soshnikov}, which study the Wigner ensemble of Hermitian/Symmetric matrices with independent
entries.

We now describe the exactly soluble model alluded to above.

\begin{definition}  A complex random variable, $X$, with values in the unit disk,
$\Disk$, is $\Theta_\nu$-distributed (for $\nu>1$) if
\begin{equation}\label{E:ThetaDefn}
\Exp\{f(X)\} = \tfrac{\nu-1}{2\pi} \int\!\!\!\int_\Disk f(z)
(1-|z|^2)^{(\nu-3)/2} \,d^2z.
\end{equation}
For $\nu\geq2$ an integer, this has the following geometric interpretation: If $v$ is chosen at random from the
unit sphere $S^\nu$ in $\Reals^{\nu+1}$ according to the usual surface measure, then $v_1+iv_2$
is $\Theta_\nu$-distributed.
\end{definition}

An elementary computation shows that if $X\sim \Theta_\nu$, then
\begin{equation}\label{ThetaMean}
\Exp\{ |X|^2 \} = \tfrac{2}{\nu+1}.
\end{equation}
The combined content of Theorem~1.2 and Proposition~B.2 from \cite{KN} is:

\begin{theorem}\label{T:KN}
Given $\beta>0$, let $\alpha_k\sim\Theta_{\beta(k+1)+1}$ be independent
random variables.  Then the eigenvalues of $\C^{(n)}$
are distributed according to the C$\beta\!$E${}_n$ ensemble.
\end{theorem}

Note that \cite{KN} and its precursor, \cite{DumE}, contain similar results that could be used in
a parallel investigation of the Jacobi matrix case.  The bulk spectral properties of one such model
are discussed from a random-matrix perspective in \cite{ForrSmil}.

The bulk of the proof of Theorem~\ref{T:main}(ii) appears in Section~\ref{S:Critical} although it also
relies on material from Section~\ref{S:Basic}.

Next we present our new representation of the C$\beta$E process.
The set of critical values of any non-decreasing function $\psi\!:\!\Reals\to\Reals$
(i.e., values taken at more than one point) is at most countable; for otherwise one could
find uncountably many disjoint open subsets of $\Reals$.
Therefore, for all but countably many $v\in(\inf \psi,\sup \psi)$ we may uniquely define $x=\psi^{-1}(v)$ as the
point where $\psi$ passes through (or jumps past) the value $v$.

\begin{definition}\label{Dinc}
Given a probability measure on non-decreasing functions $\psi\!:\!\Reals\to\Reals$, with expectation $\Exp_\psi$,
we define an associated point process by
\begin{equation}
\Exp\{X_f\} = \Exp_\psi \int_0^{2\pi} \exp\biggl\{ - \sum_{m\in\Ints} f\circ\psi^{-1}(2\pi m+\omega) \biggr\}\,d\omega,
\end{equation}
that is, points are placed at the inverse images of a random translate of $2\pi\Ints$.
\end{definition}

For example, the picket fence process is associated to $\psi(x)=x$ (non-random), while a
Poisson point process arises when $\psi$ is chosen to be ($2\pi$ times) the Poisson jump process.
In general, the average number of points in the interval $[a,b)$ is $\Exp\{\psi(b-)-\psi(a-)\}$.

The random non-decreasing function that is relevant for us is the relative Pr\"ufer phase.
This is described precisely in Section~\ref{S:Basic}; however, we would like to give some of
the flavour of this here.

The notion of Pr\"ufer phase originates in the study of Sturm--Liouville problems, see for example,
\cite[Ch. 8]{CodLev}.  Given $q\in L^1$, let us define $u(x;E)$ as the solution of
$$
-u''(x) + (q(x)-E) u(x) = 0
$$
on $[0,L]$ that obeys $u(0)=0$ and $u'(0)=1$. The Pr\"ufer phase $\psi(x;E)$ and amplitude $R(x;E)$ are defined by
$$
u(x;E) = R(x;E) \sin(\psi(x;E)),\qquad u'(x;E) = R(x;E) \cos(\psi(x;E))
$$
and $\psi(0;E)=0$.  Notice that if we place a self-adjoint boundary condition $u' \sin(\eta) - u \cos(\eta) =0$
at $x=L$, then the eigenvalues of the Sturm--Liouville problem are the energies where $\psi(L;E)=\eta$.  Note
also that $\psi(L,E)$ is a strictly increasing function of $E$; this is essentially the Sturm Oscillation Theorem.

All the properties just described carry over to the natural CMV analogue of Pr\"ufer variables (see
Section~\ref{S:Basic} or \cite[\S 10.12]{SimonOPUC2}).  In the microscopic scaling we consider, the
key quantity is the Pr\"ufer phase relative to that at a reference energy.  Because our model has rotational
symmetry, we can choose the reference energy to be $z=1$ without loss of generality.
In this way, we will find a natural process of increasing functions $\psi_k:(-\pi,\pi)\to\Reals$ so that the
eigenvalues of $\C^{(n)}$ lie at the angles where $\psi_{n-1}$ matches that random boundary condition $\eta$.
Actually, that is a lie since $\psi$ is merely the \emph{relative} Pr\"ufer phase; nevertheless, because
of the way we choose our boundary condition, it is true that the eigenvalues of $\C^{(n)}$ are related to
the random increasing function $\psi_{n-1}$ in the sense of Definition~\ref{Dinc}.

The key to proving both that the C$\beta$E${}_n$ processes converge and part (ii) of Theorem~\ref{T:main} is showing
that in the natural scaling, $\Psi_n(x) = \psi_{n-1}(\tfrac xn)$, the relative Pr\"ufer phase process converges as
$n\to\infty$.  As we will see in Section~\ref{S:Critical}, this is indeed the case; moreover we find an alternate
realization of the limiting distribution, which we now describe.

Let $B_1(t)$ and $B_2(t)$ be independent Brownian motions.  From Proposition~\ref{P:uniq}, it follows that
for each $x$, the stochastic differential equation
\begin{equation}\label{PsiSDE1}
d\Psi = x\,dt + \tfrac{2}{\sqrt{\beta t}} \Im\left\{ \bigl[ e^{i\Psi} - 1 \bigr] \bigl[dB_1(t)+idB_2(t)\bigr] \right\}
\end{equation}
has a unique solution $\Psi(t;x)$ that obeys $\Exp\{\Psi(t;x)\}=xt$ and $x\Psi(t;x)\geq 0$.  We will prove that for any
finite collection $x_1,\ldots,x_p$ the vector-valued random variable $\Psi_n(x_\mu/n)$ converges in distribution to
$\Psi(t=1;x_\mu)$.  By Proposition~\ref{P:dist_conv}, this notion of convergence is strong enough to imply that
the Laplace functionals of the eigenvalue process converge to the point process associated to the random
increasing function $\Psi(t=1,x)$.

While we where typing the last section of this work, two preprints, \cite{EdSut,RamirezRiderVirag}, appeared on the
arXiv.  These works (independent of our own) discuss matters related to our proof of part (ii) of Theorem~\ref{T:main}.
The latter preprint discusses the edge statistics for the Gaussian $\beta$ ensembles in the $n\to\infty$ limit.
Specifically, they show that the in the (naturally re-scaled) limit, the point process can be modelled as the
eigenvalues of a half-line Schr\"odinger operator with stochastic potential.  Our representation of the limiting
C$\beta$E ensemble has an analogous interpretation: Equation \eqref{PsiSDE1} describes the relative Pr\"ufer phase
for the Kre\u{\i}n system with `potential' $A(x)=(\beta x)^{-1/2}[dB_3(t)+idB_4(t)]$, where $B_3(x)$ and $B_4(x)$
are independent Brownian motions.  In this way, we may say that the C$\beta$E ensemble is modelled by the
eigenvalues of this Kre\u{\i}n system with random self-adjoint boundary condition at $x=1$.  (For a pedagogical
account of the theory of Kre\u{\i}n systems, we recommend \cite{Denisov}.)  The remarks in the final section
of \cite{EdSut} seem to indicate that the authors were seeking a result of this type.

\subsection*{Notation}  We will write $X\lesssim Y$ to indicate that $X\leq C Y$ for some constant $C$.  Typically,
this is an absolute numerical constant whose exact value is of no consequence.  When it depends some parameters, we
will endeavour to point this out.

\subsection*{Acknowledgements}
We would like to thank Marek Biskup for some probabilistic pointers. We would also like to thank Tom Spencer and
Jinho Baik for their encouragement to pursue this line of investigation.  R. K. was supported in part, by NSF grant
DMS-0401277 and a Sloan Foundation Fellowship.

\section{The basic process}\label{S:Basic}

The theory of CMV matrices is intimately connected to the theory of orthogonal polynomials on the
unit circle, just as Jacobi matrices are inherently tied to polynomials orthogonal with respect to
a probability measure on the real line.  We will confine ourselves to a review of the material we need;
a more systematic treatment can be found in \cite{SimonOPUC1}.

Let $d\mu$ denote the spectral measure associated to $\C$ (a semi-infinite CMV matrix) and the
vector $e_1=(1,0,0,\ldots)$.  As $\C$ is unitary and $e_1$ is a unit vector, this is a probability measure
on the unit circle in $\Cmplx$.  Applying the Gram--Schmidt procedure to $\{1,z,z^2,\ldots\}$ leads to a
sequence of monic orthogonal polynomials. We write $\Phi_k(z)$ for the polynomial of degree $k$.

Two important properties carry over from the Jacobi matrix case:
Firstly, the orthogonal polynomials obey a recurrence relation:
\begin{equation}
\begin{aligned}\label{SzegoRec}
\Phi_{k+1}(z) &= z\Phi_{k}(z) - \bar\alpha_k \Phi_{k}^*(z) \\
\Phi_{k+1}^*(z) &= \Phi_{k}^*(z) - \alpha_k z\Phi_{k}(z)
\end{aligned}
\end{equation}
with $\Phi_0^*=\Phi_0=1$.  Here $\Phi_k^*$ denotes the reversed polynomial,
\begin{equation}\label{PhiReflect}
\Phi_{k}^*(z) = z^k \overline{\Phi_{k}(\bar z^{-1})}.
\end{equation}
Secondly, the characteristic polynomial of finite-volume truncations can be expressed in terms of these monic
polynomials; specifically,
\begin{equation}\label{CharPoly}
\det[z-\C^{(n)}] = z\Phi_{n-1}(z) - e^{-i\eta} \Phi_{n-1}^*(z).
\end{equation}

The CMV analogue of the Weyl $m$-function used in connection with Jacobi matrices is the Carath\'eodory function,
\begin{equation}\label{E:Carat}
F(z) = \int \frac{e^{i\theta}+z}{e^{i\theta}-z}\,d\mu(\theta) = \langle e_1 | (\C+z)(\C-z)^{-1} e_1\rangle,
\end{equation}
which will be important in our treatment of part (iii) of Theorem~\ref{T:main}.  The other two parts of that
theorem rely on analysis of the relative Pr\"ufer phase, which we will now describe.

Let $B_k(z) = z \Phi_{k}(z) / \Phi_{k}^*(z)$.  In view of \eqref{PhiReflect}, this is a finite Blaschke product;
moreover, by \eqref{SzegoRec},
\begin{equation}\label{Blaschrec}
B_{k+1}(z) = z B_k(z) \frac{1-\bar\alpha_k \bar B_k(z)}{1-\alpha_k B_k(z)}.
\end{equation}
Note also that $B_0(z)=z$.  It is natural to define the (absolute) Pr\"ufer phase as the argument of $B_k(z)$.
By the Cauchy-Riemann equations, the argument of a finite Blaschke product is strictly increasing and by
\eqref{CharPoly}, the eigenvalues of $\C^{(n)}$ are precisely the points where $B_{n-1}(z)=e^{i\eta}$.

The relative Pr\"ufer phase is the sequence of random continuous increasing functions $\psi_k:(-\pi,\pi)\to\Reals$
defined by
$$
\exp\{ i\psi_k(\theta) \} = \frac{B_k(e^{i\theta})}{B_k(1)}
 = e^{i\theta} \frac{\Phi_k(e^{i\theta}) \Phi^*_k(1) }{\Phi^*_k (e^{i\theta}) \Phi_k(1)},
\qquad 0\leq k < \infty,
$$
and $\psi_k(0)=0$.  Naturally, this sequence also obeys a recurrence relation.  The relation is simpler when expressed
in terms of a new sequence of random variables $\gamma_k$ in place of the Verblunsky coefficients.  Because we have
chosen to study the rotationally invariant problem, the two systems of coefficients follow the same law:

\begin{lemma}
For each $n$, the random variables
$$
\gamma_k = B_k(1) \alpha_k,  \quad k=0,1,2,\ldots,n-2, \quad\text{and}\quad e^{-i\omega} = B_{n-1}(1) e^{i\eta}
$$
are statistically independent.  Moreover, they have the same joint distribution as $\alpha_0,\ldots,\alpha_{n-2}$
and $e^{i\eta}$.
\end{lemma}

\begin{proof}
One merely needs to make two observations.  First, the arguments of $\alpha_k$ and $e^{i\eta}$ are all statistically
independent of one another and of the moduli of the the $\alpha_k$.  Second, the arguments are all uniformly distributed
on the circle.
\end{proof}

\begin{prop}
As defined above, $\psi_k(\theta)$ obeys the following recurrence
\begin{equation}\label{psi_rec}
\psi_{k+1} = \psi_{k} + \theta + 2 \Im \log\biggl[ \frac{1-\gamma_k}{1-\gamma_k e^{i\psi_k}} \biggr],
\qquad\psi_0=\theta,
\end{equation}
where the branch of $\log$ is chosen so as to give $0$ when $\gamma_k=0$, that is, by the convergent power series
\begin{equation}\label{Log_series}
\Upsilon(\psi,\gamma) := 2 \Im \log\Bigl[ \tfrac{1-\gamma}{1-\gamma e^{i\psi}} \Bigr] =
\Im \sum_{\ell=1}^\infty \tfrac{2}{\ell} [e^{i\ell\psi}-1] \gamma^\ell
\end{equation}
Moreover, the eigenvalues of $\C^n$ are associated to $\psi_{n-1}(\theta)$
in the sense of Definition~\ref{Dinc}.
\end{prop}

\begin{proof}
The first claim is an elementary computation: from \eqref{Blaschrec},
\begin{equation}\label{Blaschrec2}
\frac{B_{k+1}(e^{i\theta})}{B_{k+1}(1)} = e^{i\theta} \frac{B_{k}(e^{i\theta})}{B_{k}(1)}\,
        \frac{1- \alpha_k B_k(1)}{1-\bar\alpha_k \bar B_k(1)}\,
        \frac{1-\bar\alpha_k \bar B_k(e^{i\theta})}{1-\alpha_k B_k(e^{i\theta})}
\end{equation}
and so
\begin{equation}\label{Blaschrec3}
e^{ i\psi_{k+1} } = e^{i\theta + i\psi_{k} }\, \frac{1-\gamma_k}{1-\bar\gamma_k}\,
        \frac{1-\bar\gamma_k e^{ - i\psi_{k} } }{1-\gamma_k e^{ i\psi_{k} }}.
\end{equation}
To finish the proof of \eqref{psi_rec}, one should note that $\zeta/\bar\zeta  = \exp\{2i\Im\log(\zeta)\}$
for any $\zeta\in\Cmplx\setminus \{0\}$.

To prove the second claim, we note that by \eqref{CharPoly}, the eigenvalues of $\C^{(n)}$
are the points $e^{i\theta}$ where $e^{i\eta} B_{n-1}(e^{i\theta}) = 1$ or equivalently, where
$\psi_{n-1}(\theta) - \omega \in 2\pi\Ints$.  Recall that $e^{-i\omega}=B_{n-1}(1) e^{i\eta}$, which is
statistically independent of $\psi_{n-1}(\theta)$.
\end{proof}

\begin{warn}
Despite the fact that $z=1$ is an eigenvalue of $\C^{(n)}$ if and only if $\omega=0$,
$\psi_{n-1}$ is not the natural the conditioned process.  Specifically, let $f$ be a continuous function on the
unit circle, then
\begin{align*}
\Exp\bigl\{ e^{-\tr f(\C^{(n)})} &\big| 1\in\sigma(\C^{(n)}) \bigr\} \\
:=&{} \lim_{\epsilon\downarrow0}  \Exp\bigl\{ e^{-\tr f(\C^{(n)})} \big|%
    \sigma(\C^{(n)})\cap e^{i[0,\epsilon]}\neq\emptyset\bigr\} \\
=&{}\lim_{\epsilon\downarrow0} \tfrac{2\pi}{n \epsilon} \Exp\biggl\{\int_0^{\psi_{n-1}(\epsilon)}
     \exp\left[-{\textstyle\sum_m f\circ\exp\bigl(i\psi_{n-1}^{-1}(2\pi m +\omega)\bigr)}\right]
     \,\tfrac{d\omega}{2\pi} \biggr\} \\
=&{} \Exp\biggl\{  \tfrac1n \psi_{n-1}'(0)
    \exp\Bigl[-{\textstyle\sum_m f\circ\exp\bigl(i\psi_{n-1}^{-1}(2\pi m)\bigr)}\Bigr] \biggr\}.
\end{align*}
Hence the probability distribution picks up an additional weight, $\frac1n \psi_{n-1}'(0)$.  It is not difficult
to derive
$$
\psi_{n-1}'(0) = 1 + \sum_{k=0}^{n-2} \prod_{l=0}^{k} \biggl( \Re\frac{1+\gamma_l}{1-\gamma_l}\biggr)
$$
from \eqref{psi_rec}.
\end{warn}

We will use the following properties of the function $\Upsilon$:

\begin{prop}\label{P:LogEst}
Given $0\leq r<1$ and $\phi,\psi\in\Reals$,
\begin{gather}
\label{MVT}
 \int_0^{2\pi} \Upsilon(\psi,re^{i\theta}) \,\tfrac{d\theta}{2\pi} = 0 \\
\label{LogEst0}
\int_0^{2\pi}  \bigl| \Upsilon(\psi,re^{i\theta}) \bigr|^2 \,\tfrac{d\theta}{2\pi} \leq  \tfrac{4\pi^2}{3}r^2 \\
\label{LogEst}
\biggl| \int_0^{2\pi}  \Upsilon(\psi,re^{i\theta}) \Upsilon(\phi,re^{i\theta})\,\frac{d\theta}{2\pi} \biggr|
 \leq  2\bigl(|\psi|+|\phi|\bigr) \log\bigl[\tfrac{1}{1-r^2}\bigr]
\end{gather}
\begin{equation}\label{LogEst2}
\begin{aligned}
\biggl| 2 r^2 \Re\bigl\{ [e^{i\psi}-1][e^{-i\phi}-1] \bigr\} -
        \int_0^{2\pi}  \Upsilon(\psi, r & e^{i\theta}) \Upsilon(\phi,re^{i\theta})\,\tfrac{d\theta}{2\pi} \biggr| \\[2mm]
&\leq  2\bigl(|\psi|+|\phi|\bigr) \, r^2 \log\bigl[\tfrac{1}{1-r^2}\bigr]
\end{aligned}
\end{equation}
and also
\begin{equation}\label{LogEst5}
 \int_0^{2\pi} \left|\log\bigl[\tfrac{1-re^{i\theta}}{1-re^{i\theta+i\psi}}\bigr]  \right|^4
        \,\frac{d\theta}{2\pi}
    \leq 16 |\psi| \log^2\bigl[1-r^2\bigr].
\end{equation}
Lastly, interpolating between \eqref{LogEst} and \eqref{LogEst5} gives
\begin{equation}\label{LogEst3}
 \int_0^{2\pi} \left| \Upsilon(\psi,re^{i\theta}) \right|^3
        \,\frac{d\theta}{2\pi} \lesssim |\psi| \log^{3/2}\bigl[\tfrac{1}{1-r^2}\bigr].
\end{equation}
\end{prop}

\begin{proof}
Equation \eqref{MVT} follows immediately from the mean value theorem for harmonic functions.

For the second, third, and fourth claims, we apply Parseval's identity to see that
\begin{align*}
\int_0^{2\pi}  \Upsilon(\psi,re^{i\theta}) & \Upsilon(\phi,re^{i\theta})\,\tfrac{d\theta}{2\pi} \\
&= \sum_{\ell=1}^\infty \tfrac{4 r^{2\ell}}{\ell^2}
        \int \Im \bigl\{ [e^{i\ell\psi}-1] e^{i\ell\theta}\bigr\}
                \Im \bigl\{ [e^{i\ell\phi}-1] e^{i\ell\theta}\bigr\} \,\tfrac{d\theta}{2\pi} \\
&= \sum_{\ell=1}^\infty \tfrac{2 r^{2\ell}}{\ell^2} \Re\bigl\{ [e^{i\ell\psi}-1][e^{-i\ell\phi}-1] \bigr\}.
\end{align*}
Inequality \eqref{LogEst0} follows by bounding this by $8 r^2 \sum \ell^{-2}$.
As $|e^{i\theta}-1|\leq |\theta|$, we have
$$
\bigl| \Re\bigl\{ [e^{i\ell\psi}-1][e^{-i\ell\phi}-1] \bigr\} \bigr| \leq \ell |\psi| + \ell |\phi|.
$$
To deduce \eqref{LogEst}, we simply substitute this estimate into the identity above and
then sum the resulting series. The inequality \eqref{LogEst2} follows from the identity above by similar reasoning.

We now turn to \eqref{LogEst5}.  Applying Parseval's identity to the square of \eqref{Log_series} gives
\begin{equation}\label{LE5a}
 \int_0^{2\pi} \left|\log\bigl[\tfrac{1-re^{i\theta}}{1-re^{i\theta+i\psi}}\bigr]  \right|^4
        \,\frac{d\theta}{2\pi}
= \sum_{q=2}^\infty r^{2q} \Biggl| \sum_{\ell=1}^{q-1} \tfrac{1}{\ell(q-\ell)}[e^{i\ell\psi}-1]
        [e^{i(q-\ell)\psi}-1]\Biggr|^2.
\end{equation}
We may estimate the sum over $\ell$ in each of two ways: First, one may bound each modulus of each term
in square brackets by $2$. Second, as the sum is symmetrical under $\ell\mapsto(q-\ell)$, we may bound it by
$$
2\sum_{\ell=1}^{\lfloor q/2\rfloor} \tfrac{|e^{i\ell\psi}-1|}{\ell}\cdot\tfrac{|e^{i(q-\ell)\psi}-1|}{(q-\ell)}
\leq 2\sum_{\ell=1}^{\lfloor q/2\rfloor} |\psi|\cdot \tfrac{4}{q} \leq 4|\psi|.
$$
Using each of these estimates for one factor of the sum in \eqref{LE5a} gives
\begin{align}
\text{LHS\eqref{LogEst5}} \leq  16 |\psi| \sum_{q=2}^\infty r^{2q} \sum_{\ell=1}^{q-1} \tfrac{1}{\ell(q-\ell)}
= 16|\psi| \log^2\bigl[\tfrac{1}{1-r^2}\bigr],
\end{align}
which completes the proof of the proposition.
\end{proof}

\begin{coro}\label{C:mean_var}  For any rotationally invariant system of parameters,
\begin{equation}\label{E:mean}
\Exp\{\psi_s(\theta)\}=(s+1)\theta,\qquad \Exp\{|\psi_s(\theta)|\}=(s+1)|\theta|,
\end{equation}
and for $k\leq m$,
\begin{align}\label{E:Var}
\Exp\bigl\{ |\psi_{m}(\tfrac{x}{n}) - \psi_{k}(\tfrac{x}{n}) - \tfrac{x(m-k)}{n} |^2 \bigr\}
&=  \sum_{s=k}^{m-1} \Exp\bigl\{ | \Upsilon(\psi_{s}(\tfrac{x}{n}),\gamma_s) |^2 \bigr\}.
\end{align}
\end{coro}

\begin{proof}
For each $s$, $\psi_s(\theta)$ is an increasing function, which vanishes at $\theta=0$.
Thus $\psi_s(\theta)$ takes only non-negative values for $\theta>0$ and non-positive values for $\theta<0$.
For this reason, the second claim follows from the first.  Both the estimates for the mean and the variance follow
quickly from the recurrence relation, \eqref{psi_rec}, and~\eqref{MVT}.
\end{proof}

As outlined in the introduction, we will deduce convergence of the point processes from that of the relative
Pr\"ufer phase.  The next result shows how this can be done.

\begin{prop}\label{P:dist_conv}
Suppose there are processes $\Psi_n(x)$, $n\in\Ints$ and $\Psi(x)$ of non-decreasing functions with two
properties: (i) for all $x\in\Reals$,
\begin{align}\label{E conv}
\lim_{n\to\infty} \Exp\bigl\{ |\Psi_n(x)| \bigr\} = \Exp\bigl\{ |\Psi(x)| \bigr\} < \infty
\end{align}
and (ii) for any finite collection of $x_j\in\Reals$, the joint distribution of
$\Psi_n(x_j)$ converges to that of $\Psi(x_j)$.  Then the point processes associated to $\Psi_n$ converge
in distribution to that associated to $\Psi$.
\end{prop}

\begin{proof}
The probabilistic portion of the proof is quite simple once we have the right deterministic information.
To this end, let $\Psi$ and $\Phi$ be non-decreasing functions on $\Reals$ and let $f\in C^\infty_c(\Reals)$
be non-negative.  Elementary considerations show that
\begin{align}\label{Lap_diff}
\int_0^{2\pi} \biggl| &\exp\biggl\{ - \sum_{m\in\Ints} f\circ\Phi^{-1}(2\pi m+\omega) \biggr\}
    - \exp\biggl\{ - \sum_{m\in\Ints} f\circ\Psi^{-1}(2\pi m+\omega) \biggr\} \biggr| \,d\omega \\
&\leq \int_0^{2\pi} \sum_{m\in\Ints}\,
    \bigl| f\circ\Phi^{-1}(2\pi m+\omega) - f\circ\Psi^{-1}(2\pi m+\omega) \bigr| \,d\omega \\
&\leq \int_\Reals \bigl| f\circ\Phi^{-1}(\omega) - f\circ\Psi^{-1}(\omega) \bigr| \,d\omega
\end{align}
Next, we choose $[a,b]\subset \Reals$ so that $\supp(f)\subset[a,b]$.  Then
$$
|f(x)-f(y)| \leq  |\tilde x - \tilde y| \cdot \|f'\|_{L^\infty}
\quad\text{where}\quad
\tilde x = \begin{cases} a &:\ x<a \\ x&:\ x\in[a,b]\\ b &:\ x>b\end{cases}
$$
and similarly for $\tilde y$.  Therefore,
\begin{align}
\int_\Reals |f\circ\Phi^{-1}(\omega) - f\circ\Psi^{-1}(\omega)| \,d\omega
&\leq \|f'\|_{L^\infty} \int_\Reals |\widetilde{\Phi^{-1}(\omega)} - \widetilde{\Psi^{-1}(\omega)}| \,d\omega \\
&= \|f'\|_{L^\infty} \int_a^b |\Phi(x) - \Psi(x)| \,dx.  \label{phipsi_diff}
\end{align}
The equality follows by realizing that both integrals give the area of the region between the
graphs of $\Phi$ and $\Psi$ over the interval $[a,b]$ (with vertical lines drawn at any jumps).

Combining our efforts so far, we have bounded the expression in \eqref{Lap_diff} by that in~\eqref{phipsi_diff}.
We will make one further reduction.  Given an integer $p\geq 1$, let us partition $[a,b]$ into $p$ equal intervals
by introducing $x_j=a+j(b-a)/p$ with $0\leq j\leq p$.  As both $\Phi$ and $\Psi$ are non-decreasing,
\begin{align}
\int_{x_j}^{x_{j+1}} &|\Phi(x) - \Psi(x)| \,dx \\
    &\leq \tfrac{b-a}{p} \bigl[ \max\{\Phi(x_{j+1}),\Psi(x_{j+1})\} - \min\{\Phi(x_j),\Psi(x_j)\} \bigr] \\
    &\leq \tfrac{b-a}{p} \bigl[ |\Psi(x_{j+1})-\Psi(x_j)| + |\Phi(x_{j+1})-\Psi(x_{j+1})|
        +  |\Phi(x_j)-\Psi(x_j)| \bigr],
\end{align}
which shows that
\begin{align}\label{disc_int_est}
\int_a^b |\Phi(x) - \Psi(x)| \,dx \leq \tfrac{b-a}{p} |\Psi(b)-\Psi(a)|
        + 2\tfrac{b-a}{p} \sum_{j=0}^p |\Phi(x_{j})-\Psi(x_{j})|.
\end{align}

Let us now begin the probabilistic portion of the argument.  We will think of $\Psi_n(x_j)$ and $\Psi(x_j)$
as vector-valued random variables where $0\leq j\leq p$ indexes the components.  By the Skorohod coupling,
we can realize the random variables $\Psi(x_j)$ and all $\Psi_n(x_j)$
on a single probability space so that $\Psi_n(x_j)\to\Psi(x_j)$ almost-surely.  Using hypothesis \eqref{E conv},
we can up-grade this to
$$
\lim_{n\to\infty} \Exp\bigl\{  |\Psi_n(x_j)-\Psi(x_j)|  \bigr\} =0,
$$
which by virtue of \eqref{disc_int_est}, implies
\begin{align}
\limsup_{n\to\infty} \int_a^b |\Psi_n(x) - \Psi(x)| \,dx \leq \tfrac{b-a}{p} \Exp\bigl\{ |\Psi(b)-\Psi(a)| \bigr\}
\end{align}
for any positive integer $p>0$.  Tracing back the deductions made earlier, we can conclude convergence of the Laplace
functionals.
\end{proof}

\noindent\textit{Remark:}  Once one knows that the distributions of $\Psi_n(x_j)$ converge for any finite set of $x_j$,
the existence of a limiting \emph{process}, $\Psi(x)$, follows from the Kolmogorov extension theorem.

\section{Fast decay}\label{S:Fast}

\begin{theorem}
If\/ $\Exp \left\{ |\alpha_s|^2 \right\}=o(s^{-1})$ then $\psi_{n-1} (\tfrac{x}{n}) \to x$ in $L^1$-sense
for each $x\in[0,\infty)$.  This implies that locally the eigenvalues follow clock statistics.
\end{theorem}

\begin{proof}
As $\Exp\left\{ |\gamma_s|^2 \right\}= o(s^{-1})$,
$$
\sum_{n\geq s \geq \epsilon n} \Exp\left\{ |\gamma_s|^2 \right\} = o(1)
$$
as $n\to\infty$ for any choice of $\epsilon$.  Therefore, there is a sequence $k_n$ that is $o(n)$
and obeys
$$
\sum_{s=k_n}^n \Exp\left\{ |\gamma_s|^2 \right\} = o(1)
$$
as $n\to\infty$. Combining this with \eqref{E:Var} and \eqref{LogEst0} we have,
$$
\Exp\bigl\{ |\psi_{n-1}(\tfrac{x}{n}) - \psi_{k_n}(\tfrac{x}{n}) - (n-1-k_n)\tfrac{x}{n} |^2 \bigr\}
\leq \tfrac{4\pi^2}{3} \sum_{s=k_n}^{n-2}\Exp\{ |\gamma_s|^2 \} = o(1).
$$
Now by the triangle inequality and the Cauchy--Schwarz inequality,
\begin{align*}
\Exp\{ |\psi_{n-1} (\tfrac{x}{n}) - x| \}
    &\leq \Exp\{ |\psi_{k_n} (\tfrac{x}{n}) - (k_n+1)\tfrac{x}{n}| \}
    + o(1) \\
&\leq  2(k_n+1)\tfrac{|x|}{n}  + o(1),
\end{align*}
which is $o(1)$ by the way we chose $k_n$.  Note that the second inequality follows from~\eqref{E:mean}.

Convergence of the point processes follows from Proposition~\ref{P:dist_conv}.
\end{proof}

\noindent\textit{Remark.} If $\Exp \left\{ |\alpha_k|^2 \right\} \in \ell^1$,
then one obtains clock spacing on an almost macroscopic scale.  Indeed for any (non-random) sequence
$x_n$ with $|x_n|=o(n)$,
$$
  \Exp\bigl\{ |\psi_{n-1}(x_n/n)  - x_n| \bigr\} \to 0
$$
as $n\to\infty$.  The proof follows the outline above, but with $k_n\to\infty$ chosen so that $x_nk_n=o(n)$.

\section{Critical decay}\label{S:Critical}

Given $\vec x\in\Reals^p$ with coordinates $x_\mu$, we define a sequence of $\Reals^p$-valued
processes $\Psi_n(t)$ on $[0,\infty)$ as follows: for all $1\leq j \leq p$,
\begin{align}
\Psi_{n,\mu}(0)&=0 \\
\Psi_{n,\mu} \bigl(\tfrac{k+1}{n}\bigr) &= \psi_k(\tfrac{x_\mu}n)\qquad\text{for $0\leq k < \infty$}
\end{align}
and for intermediate values of $t$, we define $\Psi_n(t)$ by linear interpolation.  The goal of this section
is to prove that this sequence of processes converge to a limit that is independent of the specifics of the
distribution of the random variables $\gamma_k$. Our first job will be to prove the existence of even just a
subsequential limit of these processes.

Consistent with our goal of proving part (ii) of Theorem~\ref{T:main}, will we assume for the remainder of
this section that
\begin{align}
\Exp\left\{ |\gamma_k|^2 \right\} &= \tfrac{2}{\beta(k+1)}(1 + \e_k )        \label{E:Ass1}\\
\Exp\left\{ \log^{2}\Bigl[\tfrac1{1-|\gamma_k|^2}\Bigr] \right\} &\leq \tfrac{1}{(k+1)} \e_k   \label{E:Ass2}\\
\Exp\left\{ \log^{3/2}\Bigl[\tfrac1{1-|\gamma_k|^2}\Bigr] \right\} &\leq \tfrac{1}{(k+1)} \e_k \label{E:Ass3a}
\end{align}
for some decreasing sequence of positive numbers $\e_k$ converging to zero.  Note that while \eqref{E:Ass3a} does
not appear among the assumptions of this part of the Theorem, it is easily deduced from them.  Similarly,
\begin{align}   \label{E:Ass3b}
\Exp\left\{ \log\Bigl[\tfrac1{1-|\gamma_k|^2}\Bigr] \right\} &\lesssim  \tfrac{1}{(k+1)}.
\end{align}
Here and for the remainder of this section the implicit constants hidden in the $\lesssim$ notation
are permitted to depend on $\beta$, the properties of the sequence $\e_k$ and on the values of $x$
under consideration; they will never depend on $n$.

Before proceeding, let us pause to demonstrate that the exactly soluble model of Theorem~\ref{T:KN} obeys
the hypotheses given above.  If $\alpha\sim\Theta_\nu$, then
\begin{align*}
\Exp\Big\{ \log^m\Bigl[ \tfrac{1}{1-|\alpha|^2}\Bigr] \Big\}
&= \tfrac{\nu-1}{2} \int_0^\infty t^m e^{-(\nu-1)t/2} \,dt
= \Gamma(m+1) \left(\tfrac{2}{\nu-1}\right)^{m},
\end{align*}
by making the change of variables $e^{-t}=1-|\alpha|^2$.

\begin{lemma}\label{L:tight}
For fixed $x$,
\begin{align}\label{E:square diff}
\Exp\bigl\{ |\psi_{m}(\tfrac{x}{n}) - \psi_{k}(\tfrac{x}{n}) - \tfrac{x(m-k)}{n} |^2 \bigr\}
&\lesssim \tfrac{m-k}{n}.
\end{align}
In particular, from $k=0$ we have $\Exp\bigl\{ |\psi_{s}(\tfrac{x}{n})|^2 \bigr\}
\lesssim \tfrac{s}{n} + (\tfrac{s+1}{n})^2$. Furthermore,
\begin{align}\label{E:four diff}
\Prob\Bigl\{ \sup_{k\leq s\leq m} |\psi_{m}(\tfrac{x}{n}) - \psi_{k}(\tfrac{x}{n}) - \tfrac{x(m-k)}{n} |
        \geq \epsilon \Bigr\}
&\lesssim \epsilon^{-4} \bigl( \tfrac{m-k}{n}\log\bigl[\tfrac{m+1}{k+1}\bigr] + \tfrac{m-k}{n} \e_k \bigr)
\end{align}
\end{lemma}

\begin{proof}
Let $k$ be fixed and define
$$
\phi_s=\psi_{s}(\tfrac{x}{n}) - \psi_{k}(\tfrac{x}{n}) - \tfrac{x(s-k)}{n}
$$
for $s\geq k$.  Note that $\phi_{s+1}=\phi_s+\Upsilon(\psi_s,\gamma_s)$.  Equation \eqref{MVT} shows that
this is a martingale; therefore, using \eqref{LogEst}, \eqref{E:Ass3b} and \eqref{E:mean}
\begin{align*}
\Exp\bigl\{ |\phi_{m}|^2 \bigr\} &= \sum_{s=k}^{m-1} \Exp\bigl\{ |\Upsilon(\psi_s,\gamma_s)|^2 \bigr\}
\lesssim \sum_{s=k}^{m-1} \Exp\Bigl\{ |\psi_s| \log\Bigl[\tfrac1{1-|\gamma_s|^2}\Bigr] \Bigr\}
\lesssim \tfrac{m-k}{n}.
\end{align*}
This proves \eqref{E:square diff}; we now consider \eqref{E:four diff}.  By \eqref{MVT},
\begin{equation}\label{fourth power}
\begin{aligned}
\Exp\bigl\{ |\phi_{s+1}|^4 \bigr\} =
\Exp\bigl\{ |\phi_{s}|^4 \bigr\} & + 6\Exp\bigl\{ |\phi_{s}|^2 |\Upsilon(\psi_s,\gamma_s)|^2 \bigr\} \\
&+ 4\Exp\bigl\{ \phi_{s} \Upsilon(\psi_s,\gamma_s)^3 \bigr\} + \Exp\bigl\{ |\Upsilon(\psi_s,\gamma_s)|^4 \bigr\}.
\end{aligned}
\end{equation}
Using \eqref{LogEst0} and \eqref{E:square diff},
\begin{align*}
\Exp\bigl\{ |\phi_{s}|^2 |\Upsilon(\psi_s,\gamma_s)|^2 \bigr\} \lesssim \Exp\bigl\{ |\phi_{s}|^2 |\gamma_s|^2 \bigr\}
\lesssim \tfrac{s-k}{n}\tfrac{1}{s+1}
\end{align*}
while by \eqref{LogEst3}, \eqref{E:mean} and \eqref{E:Ass3a},
\begin{align*}
\Exp\bigl\{ |\phi_{s}| |\Upsilon(\psi_s,\gamma_s)|^3 \bigr\}
&\lesssim \Exp\Bigl\{ \bigl[|\psi_{s}|+|\psi_k| + (s-k)\tfrac{x}{n}\bigr]
        \log^{3/2}\Bigl[\tfrac1{1-|\gamma_s|^2}\Bigr] \Bigr\} \\
&\lesssim \tfrac{s+1}{n} \tfrac{1}{s+1} \e_s .
\end{align*}
Similarly using \eqref{LogEst5} and \eqref{E:Ass2},
\begin{align*}
\Exp\bigl\{ |\Upsilon(\psi_s,\gamma_s)|^4 \bigr\} \lesssim \tfrac{s+1}{n} \tfrac{1}{s+1} \e_s.
\end{align*}
By substituting the last three estimates into \eqref{fourth power}, we obtain
\begin{align*}
\Bigl| \Exp\bigl\{ |\phi_{s+1}|^4 \bigr\} - \Exp\bigl\{ |\phi_{s}|^4 \bigr\} \Bigr|
&\lesssim \tfrac{s-k}{n}\tfrac{1}{s+1} + \tfrac{1}{n} \e_s
\end{align*}

By summing the preceding estimate over $s$ and using the fact that $\phi_k\equiv 0$, we obtain
\begin{align*}
\Exp\bigl\{ |\phi_m|^4 \bigr\} & \lesssim \tfrac{m-k}{n}\log\bigl[\tfrac{m+1}{k+1}\bigr] + \tfrac{m-k}{n} \e_k.
\end{align*}
Thus \eqref{E:four diff} follows by the martingale maximal theorem.
\end{proof}

\begin{prop} For fixed $\vec x$, the processes $\Psi_n(t)$ define a tight family of measures
on $C([0,\infty);\Reals^p)$.
\end{prop}

\begin{proof}
We will prove that for fixed $x$ and $T\geq 1$,
$$
\limsup_{n\to\infty} \Prob\Bigl( 2\epsilon \leq \sup\bigl\{
    | \psi_s\bigl(\tfrac{x}{n}\bigr) - \psi_t\bigl(\tfrac{x}{n}\bigr) | :
    0\leq s < t \leq nT \text{ and } |s-t|\leq n \delta \bigr\} \Bigr)
=o(1)
$$
as $\delta\downarrow 0$.  The Ascoli--Arzela theorem shows that this suffices; cf. \cite[\S 1.4]{SV}.

It suffices to consider the case $n\delta > 2$.
We wish to break the problem into blocks; to this end, let $k_j=\lceil jn\delta \rceil$, that is,
$jn\delta$ rounded up to an integer.  The zeroth block is easiest: as $\psi_s(\tfrac{x}{n})-(s+1)\tfrac{x}n$
is a martingale,
\begin{equation}\label{zeroblock}\begin{aligned}
\Prob\Bigl\{ | \psi_s\bigl(\tfrac{x}{n}\bigr) | \geq \epsilon \text{ for some } 0\leq s \leq k_1 \Bigr\}
&\lesssim \epsilon^{-1} \Exp\bigl\{ |\psi_{k_1}(\tfrac{x}{n})|+(k_1+1)\tfrac{|x|}{n} \bigr\} \\
&\lesssim \delta \epsilon^{-1}
\end{aligned}\end{equation}
Naturally this bounds the probability that $| \psi_s - \psi_t | \geq 2\epsilon$ for any pair $s,t$ in the interval
$[0,k_1]$.

For $j\geq1$, we can apply Lemma~\ref{L:tight} with $\epsilon$ half the size and so obtain
\begin{align*}
\Prob\Bigl\{ \bigl|\psi_s&\bigl(\tfrac{x}{n}\bigr)-\psi_t\bigl(\tfrac{x}{n}\bigr) \bigr| \geq \epsilon
        \text{ for some } k_j \leq s < t \leq k_{j+1} \Bigr\} \\
&\lesssim \epsilon^{-4} \Bigl[ \delta \log\bigl[\tfrac{j}{j+1}\bigr] + \delta \e_{k_j} + (\delta x)^4\Bigr].
\end{align*}
Because $\e_s$ decreases to zero,
$$
J^{-1} \sum_{j=1}^J \e_{k_j} \leq J^{-1} \sum_{j=1}^J \e_{j} = o(1)
$$
as $J\to\infty$ uniformly in the region $n\delta > 2$. Combining this with $\log[j/(j+1)]\lesssim j^{-1}$,
we obtain
\begin{equation*}\begin{aligned}
\limsup_{n\to\infty} \sum_{j=1}^{\lceil T/\delta \rceil}
    \Prob&\Bigl\{ |\psi_s\bigl(\tfrac{x}{n}\bigr)-\psi_t\bigl(\tfrac{x}{n}\bigr) | \geq \epsilon
    \text{ for some } k_j \leq s < t \leq k_{j+1} \Bigr\} \\
&\lesssim \epsilon^{-4} \Bigl[ \delta \log\bigl[1+T\delta^{-1}\bigr] + o(1) + \delta^3\Bigr] = o(1)
\end{aligned}\end{equation*}
where $o(1)$ means as $\delta\downarrow0$ for $T$ fixed.

Combining the estimate just proved with \eqref{zeroblock} gives the claim made in the first paragraph of the proof.
Note that the change from $\epsilon$ to $2\epsilon$ comes from the fact that the points $s$ and $t$ may lie in
adjacent blocks rather than the same one.
\end{proof}

We now begin to examine the properties of any subsequential limit of $\Psi_n(t)$; specifically, we will prove that
any such limit is the solution of a certain martingale problem.  This approach to studying diffusions was introduced
by Stroock and Varadhan and is very convenient for proving convergence; indeed much of this section modelled
on \cite[\S11.2]{SV}.

We can no longer treat the different values of $x$ separately.  For this reason, it is helpful to
introduce some notation for the discrete vector problem.  Because the equations are going become rather long, we
need to make it as compact as possible.  Let $\vpn_s$ denote the vector with components
$$
\vpn_{s,\mu} = \psi_s(\tfrac{x_\mu}n),\qquad \mu=1,2,\ldots,p.
$$
Our first lemma shows that this discrete process almost solves a martingale problem.

\begin{lemma}\label{L:discrete martingale}
Given $\vec x$ and $f\in C^\infty_c(\Reals^p)$, let
$$
\tilde M_n(m,k) = f(\vpn_m) - \sum_{s=k}^{m-1} \biggl[ x_\mu f_\mu (\vpn_s)
        + \tfrac{2}{\beta(s+1)} f_{\mu\nu}(\vpn_s) \Re\bigl\{ [e^{i\vpn_{s,\mu}}-1][e^{-i\vpn_{s,\nu}}-1] \bigr\} \biggr]
$$
where subscripts on $f$ denote partial derivatives and $\mu,\nu$ are summed over whenever repeated.
Then for fixed integer $T>0$,
$$
\sup_{0\leq k<m \leq Tn} \Bigl| \Exp\bigl\{ \tilde M_n(m,k) - f(\vpn_{k}) \big| \vpn_k \bigr\}  \Bigr| \to 0
$$
in $L^1$ sense as $n\to\infty$.
\end{lemma}

\begin{proof}
The recurrence relation for $\psi$ says
\begin{align}
\vpn_{s+1,\mu} &= \vpn_{s,\mu} + \tfrac{x_\mu}n + \Upsilon\bigl(\vpn_{s,\mu},\gamma_s\bigr).
\end{align}
Using this we expand $f(\vpn_{s+1})$ in a Taylor series to third order.  By incorporating \eqref{MVT} we find
\begin{align}\label{delta f}
\Exp\bigl\{ f(\vpn_{s+1}) \big| \vpn_s \bigr\}
&= f(\vpn_s) + \tfrac{x_\mu}n f_\mu(\vpn_s) + A_s^{\mu\nu} f_{\mu\nu}(\vpn_s) + b_s
\end{align}
where $A_s$ is the sequence of matrices
$$
A_s^{\mu\nu} = \tfrac{x_\mu x_\nu}{2n^2}
    + \Exp \Bigl\{ \tfrac12 \Upsilon\bigl(\vpn_{s,\mu},\gamma_s\bigr)
            \Upsilon\bigl(\vpn_{s,\nu},\gamma_s\bigr) \Big| \vpn_s \Bigr\}
$$
and $b_s$ are scalars obeying
$$
|b_s| \lesssim \tfrac{1}{n^3} + \sup_\mu \Exp \Bigl\{
        \bigl| \Upsilon\bigl(\vpn_{s,\mu},\gamma_s\bigr) \bigr|^3 \Big| \vpn_s \Bigr\}.
$$
(The implicit constant here depends on $\vec{x}$ and the properties of $f$; this will be the case throughout
the proof.) These expressions can be better understood using results from Proposition~\ref{P:LogEst}.
By \eqref{LogEst2}, we have
\begin{align*}
\biggl| \Exp \Bigl\{ \tfrac12 \Upsilon\bigl(\vpn_{s,\mu},\gamma_s\bigr)&
        \Upsilon\bigl(\vpn_{s,\nu},\gamma_s\bigr) \Big| \vpn_s \Bigr\}
- \Exp\bigl\{|\gamma_s|^2\bigr\}\Re\bigl\{ [e^{i\vpn_{s,\mu}}-1][e^{-i\vpn_{s,\nu}}-1] \bigr\}  \biggr| \\
&\lesssim |\vpn_{s,\mu}+\vpn_{s,\nu}| \Exp\Bigl\{ |\gamma_s|^2 \log\Bigl[ \tfrac1{1-|\gamma_s|^2} \Bigr] \Bigr\} \\
&\lesssim |\vpn_{s,\mu}+\vpn_{s,\nu}| \Exp\Bigl\{ \log^{3/2}\Bigl[ \tfrac1{1-|\gamma_s|^2} \Bigr] \Bigr\}
\end{align*}
and so by invoking \eqref{E:mean} and \eqref{E:Ass3a} we obtain
\begin{equation}\label{c_bound}
\sum_{s=0}^{nT} \Exp\Bigl\{ \Bigl| A_s^{\mu\nu} - \tfrac{2}{\beta(s+1)}
    \Re\bigl\{ [e^{i\vpn_{s,\mu}}-1][e^{-i\vpn_{s,\nu}}-1] \bigr\}\Bigr| \Bigr\}
\lesssim \tfrac Tn + \tfrac1n \sum_{s=0}^{nT} \e_s = o(1)
\end{equation}
as $n\to\infty$ with $T$ fixed.  Similarly by \eqref{LogEst3}, we have
$$
\Exp \Bigl\{ \bigl| \Upsilon\bigl(\vpn_{s,\mu},\gamma_s\bigr) \bigr|^3 \Big| \psi_s \Bigr\}
\lesssim |\vpn_{s,\mu}| \Exp\Bigl\{ \log^{3/2}\Bigl[\tfrac1{1-|\gamma_s|^2}\Bigr] \Bigl\}
$$
and so by \eqref{E:Ass2},
\begin{equation}\label{d_bound}
\sum_{s=0}^{nT} \Exp\{ |b_s| \} =o(1)
\end{equation}
as $n\to\infty$.

Combining \eqref{delta f}, \eqref{c_bound}, and \eqref{d_bound}, we may conclude that
\begin{align}\label{E:dmme}
\lim_{n\to\infty} \sum_{s=0}^{nT} \Exp\left\{\left|
        \Exp\bigl\{ \tilde M_n(s+1,s) - f(\vpn_s) \,\big| \vpn_s \bigr\} \right|\right\} =0,
\end{align}
which is the key estimate for the completion of the proof.

For any $k<m$,
\begin{align*}
\Exp\bigl\{ \tilde M_n(m,k) - f(\vpn_{k}) \big| \vpn_k \bigr\}
&= \sum_{s=k}^{m-1} \Exp\Bigl\{ \tilde M_n(s+1,s) - f(\vpn_s) \Big| \vpn_k \Bigr\} \\
&= \sum_{s=k}^{m-1} \Exp\Bigl\{ \Exp\bigl\{ \tilde M_n(s+1,s) - f(\vpn_s)\,\big| \vpn_s \bigr\}  \Big| \vpn_k \Bigr\},
\end{align*}
because the process is Markovian.  Therefore,
\begin{align*}
\Exp\biggl\{ \sup_{k<m} \Bigl| \Exp\bigl\{ \tilde M_n(m,k) - f(\vpn_{k}) \big| \vpn_k \bigr\} \Bigr| \biggr\}
&\leq  \sum_{s=0}^{nT} \Exp\Bigl\{ \Bigl| \Exp\bigl\{ \tilde M(s+1,s) - f(\vpn_s)\,\big| \vpn_s \bigr\}  \Big| \Bigr\},
\end{align*}
which converges to $0$ by \eqref{E:dmme}.
\end{proof}

\begin{prop}\label{P:mart}
Let $\vec x$ be fixed and let $\Psi(t)$ be a subsequential limit of the processes $\Psi_n(t)$.  Then\\
\hbox to 1.3em{\rm (a)\hss} If $x_\mu > x_\nu$ then $\Psi_\mu-\Psi_\nu$ is a
    non-negative function with probability one.  Similarly, for each $\mu$,
    $x_\mu \Psi_\mu$ is a non-negative function with probability one. \\
\hbox to 1.3em{\rm (b)\hss} $\Exp\{ \|\Psi(t)\|^2 \} \lesssim t + t^2$ for all $t\in[0,\infty)$.\\
\hbox to 1.3em{\rm (c)\hss} $\Exp\{ \Psi_\mu(t) \} =x_\mu t$ for all $t\in[0,\infty)$.\\
\hbox to 1.3em{\rm (d)\hss} For all $t_0>0$ and any $f\in C^\infty_c(\Reals^p)$,
\begin{equation}\label{Mdefn}
\begin{aligned}
M(t) := f(\Psi(t)) &- \int_{t_0}^t x_\mu f_\mu(\Psi(\tau)) \,d\tau \\
    &- 2 \int_{t_0}^t f_{\mu\nu}(\Psi(\tau))\Re\bigl\{ [e^{i{\Psi_\mu(\tau)}}-1][e^{-i\Psi_\nu(\tau)}-1] \bigr\}
            \frac{d\tau}{\beta\tau}
\end{aligned}
\end{equation}
is an $\mathcal{M}_t$-martingale starting at $t_0$.  Here $\mathcal{M}_t$ is the sigma algebra
generated by $\Psi(\tau)$ over the interval $t_0\leq\tau\leq t$.
\end{prop}

\begin{proof}
To avoid unsightly subscripts in what follows, $n$ is restricted throughout to lie in the subsequence
that converges rather than be an arbitrary positive integer.

(a) The subset $K$ of functions with $\Psi_\mu-\Psi_\nu\geq 0$ is closed in $C([0,\infty);\Reals^p)$; therefore,
$\Prob(\Psi\in K) \geq \limsup \Prob(\Psi_n\in K) = 1$.  The same argument applies for the second statement.

(b) By taking $k=0$ in \eqref{E:square diff} from Lemma~\ref{L:tight},
$$
\Exp\bigl\{|\Psi_{n,\mu}(\tfrac{s+1}n)|^2\bigr\} =
\Exp\bigl\{ \bigl|\psi_s(\tfrac {x_\mu} n)\bigr|^2 \bigr\} \lesssim \tfrac sn + \bigl(\tfrac{s+1}n\bigr)^2.
$$
One now simply sums over $\mu$ and takes $n\to\infty$.

(c) By \eqref{E:mean}, we know that $\Exp\{ \Psi_{n,\mu}(t) \}=x_\mu t$.  Moreover by hypothesis, we know that
$\Exp\{f(\Psi_n(t)) \} \to \Exp\{f(\Psi(t)) \}$ for any bounded continuous function $f$.
The result now follows by a simple approximation argument using part (b).

(d) It suffices to show that
$$
\Exp\left\{ G\bigl(\Psi|_{[t_0,t_1]}\bigr) \bigl[ M(t) - M(t_1)\bigr] \right\} = 0
$$
for any bounded continuous function $G$ from $C([t_0,t_1];\Reals^p)$ to $[0,1]$ and any $t>t_1\geq t_0$.
As this random variable is a continuous function of $\Psi$, in the $C([t_0,t];\Reals^p)$ topology, we need
only prove that
\begin{equation}\label{weak_condit}
\lim_{n\to\infty} \Exp\left\{ G\bigl(\Psi_n|_{[t_0,t_1]}\bigr) \bigl[ M_n(t) - M_n(t_1)\bigr] \right\} = 0,
\end{equation}
where $M_n(t)$ is the analogue of \eqref{Mdefn} with $\Psi_n$ replacing $\Psi$.  This will follow from
Lemma~\ref{L:discrete martingale} once we relate it back to the discrete process. We will use the notation
$$
k = \lceil nt_1\rceil-1 \quad\text{and}\quad m = \lfloor nt\rfloor - 1.
$$
Note that $\Psi_n(t) = (1-\theta)\vpn_{m} + \theta\vpn_{m+1}$ for some $\theta\in[0,1)$, while
$\Psi_n(t_1)$ is a combination of $\vpn_{k}$ and $\vpn_{k-1}$.

The standard estimate for the trapezoid rule, \cite[\S15.19]{Apostol2}, tells us that
$$
\biggl|\tfrac{1}{2}\bigl[g(a)+g(b)\bigr] - \int_0^1 g\bigl( (1-u)a + ub\bigr)\,du\biggl|
\leq \tfrac1{12}\|b-a\|^2 \|g_{\mu\nu}\|_{L^\infty}
$$
for any $a,b\in\Reals^p$ and any $g\in C^\infty_c(\Reals^p)$.  Consequently,
\begin{align}\label{E:Trap}
\Exp\biggl\{\biggl| &\int_{t_1}^{t} g\bigl(\Psi_n(t)\bigr)\,dt
    - \frac{1}{n}\sum_{s=k}^{m} g\bigl(\vpn_s\bigr)\biggr|\biggr\} \\
        &\lesssim \frac{\|g\|_{L^\infty}}{n} \label{E:Trap2}
                + \frac{\|g_{\mu\nu}\|_{L^\infty}}{n} \sum_{s=k}^{m}  \Exp\bigl\{\|\vpn_{s+1}-\vpn_{s}\|^2\bigr\} \\
        &\lesssim \frac{\|g\|_{L^\infty}}{n} + \frac{\|g_{\mu\nu}\|_{L^\infty}}{n}. \label{E:Trap3}
\end{align}
(Here and for the remainder of the proof, implicit constants are permitted to depend on $t$ and $t_1$.)
The first term in \eqref{E:Trap2} comes from the ends of the interval.  To pass to \eqref{E:Trap3},
we used \eqref{E:square diff}:
$$
\Exp\bigl\{\|\vpn_{s+1}-\vpn_{s}\|^2\bigr\} \lesssim \tfrac1n.
$$

From the inequality immediately above we have
$$
\Exp\Bigl\{ \bigl|\Psi_n(t)-\vpn_{m}\bigr|^2 \Bigr\} +
\Exp\Bigl\{ \bigl|\Psi_n(t_1)-\vpn_{k}\bigr|^2 \Bigr\}  \lesssim \tfrac1n.
$$
Combining this with the estimate above for \eqref{E:Trap}, gives
\begin{equation}\label{E:yawn}
\Exp \bigl\{ \bigl| M_n(t)- M_n(t_1) - \tilde M_n(m,k) + f(\vpn_{k}) \bigr| \bigr\} \to 0
\end{equation}
as $n\to\infty$.  In this way, we have reduced the proof to showing that
$$
\lim_{n\to\infty} \Exp \bigl\{ G\bigl(\Psi_n|_{[t_0,t_1]}\bigr)\bigl[\tilde M_n(m,k) - f(\vpn_{k}) \bigr] \bigr\}=0.
$$

Now $G\bigl(\Psi_n|_{[t_0,t_1]}\bigr)$ is a measurable function of
$\gamma_0,\gamma_1,\ldots,\gamma_{k-1}$, while $\tilde M(m,k)$ is a function of
$\vpn_{k},\gamma_k,\ldots,\gamma_{m-1}$.  Thus the statement above follows from
$$
\lim_{n\to\infty} \Exp \bigl\{ \tilde M_n(m,k) - f(\vpn_{k}) \bigr| \vpn_{k} \bigr\} = 0,
$$
which is a consequence of Lemma~\ref{L:discrete martingale}.
\end{proof}

Naturally, the key to proving part (ii) of Theorem~\ref{T:main} is to show that (some subset of) these
properties uniquely determines the process $\Psi(t)$.  For this not only shows that the limit exists
without passing to a subsequence, but also that the limit depends only on the assumptions \eqref{E:Ass1}
and \eqref{E:Ass2}, not on the specifics of the distribution.  Because these assumptions hold for the exactly
soluble model given in Theorem~\ref{T:KN}, we may also identify the limit as C$\beta$E.

As described in \cite[\S8.1]{SV}, for $t_0>0$, any solution of the martingale problem \eqref{Mdefn}
on $[t_0,\infty)$ gives rise to a solution of the system of stochastic differential equations
\begin{equation}\label{PsiSDE}
d\Psi_\mu = x_\mu\,dt + \tfrac{2}{\sqrt{\beta t}} \Im\left\{ \bigl[ e^{i\Psi_\mu} - 1 \bigr]
    \bigl[dB_1(t)+idB_2(t)\bigr] \right\}
\end{equation}
with independent Brownian motions $B_1$ and $B_2$.  The restriction to $[t_0,\infty)$ is necessary because
we only know this result for SDEs with bounded coefficients.  As the same probability distribution gives
a solution of \eqref{PsiSDE} for all $t_0>0$, we can speak of this as a solution of \eqref{PsiSDE} on
the \emph{open} interval $(0,\infty)$.

\begin{prop}\label{P:uniq}
For each $\vec{x}\in\Reals^p$, there is exactly one non-anticipating solution of \eqref{PsiSDE} on $(0,\infty)$
that obeys $x_\mu \Psi_\mu(t) \geq 0$ for all $t>0$ and
\begin{equation}\label{E:Puniq}
    \Exp\{ \Psi_\mu(t) \} = x_\mu t.
\end{equation}
Moreover, it is a strong solution, that is, $\Psi$ is measurable with respect to the filtration generated by
the Brownian motions {\rm(}cf. \cite[\S5.2]{KaratzasShreve}{\rm)}.
\end{prop}

\begin{proof}
Existence of a solution follows from Proposition~\ref{P:mart} and \cite[\S8.1]{SV}
or \cite[\S5.4.B]{KaratzasShreve}.  To prove the other claims, we will construct a sequence
of processes $\Phi_n(t)$ on the same probability space so that $\Phi_n$ is a measurable (and non-anticipating)
with respect to the filtration generated by the Brownian motions and so that
\begin{equation}\label{E:PhiConv}
\sup\bigl\{ |\Phi_n(t)-\Psi(t)| : n^{-1}<t<n \bigr\} \to 0
\end{equation}
in probability as $n\to\infty$.

By the standard existence and uniqueness theory for SDE with Lipshitz coefficients (e.g. \cite[\S5.1]{SV} or
\cite[\S5.2.B]{KaratzasShreve}), there is a unique solution $\Phi_n(t)$ of \eqref{PsiSDE}
on the interval $[n^{-1},\infty)$ with initial data $\Phi_{n,\mu}(n^{-1})=x_\mu n^{-1}$;
moreover, $\Phi_n$ is has the measurability properties just described and
$\Exp\{\Phi_{n,\mu}(t)\} = x_\mu t$.

The process $\Psi(t)-\Phi_n(t)$ is a martingale for $t\in[n^{-1},\infty)$.
By considering separately the cases when $\Psi(n^{-1})-\Phi_n(n^{-1})$ is non-negative or non-positive
and using the fact that solutions of a SDE cannot cross (cf. \cite[p. 293]{KaratzasShreve}), it follows that
$$
\Exp\{ |\Psi_\mu(n)-\Phi_{n,\mu}(n)| \} = \Exp\{ |\Psi_\mu(n^{-1})-\Phi_{n,\mu}(n^{-1})| \}.
$$
Using the martingale property again and \eqref{E:Puniq} we obtain
$$
\Prob\Bigl\{ \sup_{n^{-1}<t<n} |\Psi(t)-\Psi_n(t)| > \epsilon \Bigr\}
\lesssim \frac{1}{\epsilon} \Exp\{ |\Psi_\mu(n^{-1})| + |\Phi_{n,\mu}(n^{-1})| \}
\lesssim \frac{1}{n\epsilon},
$$
which proves the claim.
\end{proof}

\noindent
\textit{Remarks:}  Elementary manipulations of \eqref{PsiSDE} allow one to deduce the following properties:

i) For $\lambda>0$, $\Psi(\lambda t;x)$ and $\Psi(t;\lambda x)$ follow the same law.

ii) For any fixed $a\in\Reals$, $\Psi(x,t)-\Psi(a,t)$ and $\Psi(x-a,t)$ have the same distribution.

\begin{theorem}
For any system of parameters obeying $\Exp\left\{ |\alpha_k|^2 \right\} = \tfrac{2}{\beta(k+1)} + o(k^{-1})$
and $\Exp\left\{ \log^{2}[1-|\alpha_k|^2] \right\} =o( k^{-1} )$ the eigenvalue process converges in distribution;
moreover, the limit depends only on $\beta$.
\end{theorem}

\begin{proof}
By Proposition~\ref{P:uniq}, there is a unique process with the properties described in Proposition~\ref{P:mart}.
Thus all subsequential limits of $\Psi_n(t)$ are identical and so the limit exists without passing to a subsequence.
Moreover, the interpretation given in Proposition~\ref{P:uniq} shows that the limiting distributions are consistent
under extension of the vector $\vec{x}$ (in the sense of Kolmogorov's Theorem) and so may be regarded as the marginals
of a probability measure defining a non-decreasing-function-valued process, $\Psi(x,t)$.
More importantly, Proposition~\ref{P:uniq} shows that the distribution of the limit, $\Psi(x,t)$, depends only on
$\beta$, not on the specifics of the distribution of the Verblunsky coefficients.

Tracing our definitions back, we see that $\psi_{n-1}(x/n)$ converges in distribution to $\Psi(x,t=1)$ at least for
any finite collection points $x\in\Reals$.  Therefore, by Proposition~\ref{P:dist_conv}, the eigenvalue process converges
to that associated to $\Psi(x,t=1)$ in the sense of Definition~\ref{Dinc}.  In view of the exactly soluble example
described in Theorem~\ref{T:main}, we feel justified in calling this the C$\beta$E process.
\end{proof}

\section{Slow decay}\label{S:Slow}

In this section we prove part (iii) of Theorem~\ref{T:main}.  That is, we prove that in the
$n\to\infty$ limit, the re-scaled eigenvalues behave as a Poisson point process under the assumptions
that the Verblunsky coefficients are independent, have rotationally invariant laws,
\begin{align}\label{IIIass1}
\Exp\bigl\{ (1-|\alpha_k|^2)^{-s_0} \bigr\} \lesssim 1
\end{align}
for some $s_0>0$, and
\begin{align}\label{IIIass2}
\Exp\bigl\{ |\alpha_k|^2 \bigr\} \gtrsim (1+k)^{\epsilon - 1}.
\end{align}
As in the previous sections, the constants implicit in the $\lesssim$ notation are permitted to depend
on the constants in these inequalities, $s_0$ and $\epsilon$ included.

Our goal here is to demonstrate the applicability of the methods of \cite{StoiciuPhD,MSjat}, which prove this
result when the Verblunsky coefficients $\alpha_k$ are identically distributed (and so non-decaying).
The approach taken there is in turn derived from the work of Minami, \cite{Minami}, on eigenvalue statistics
for the Anderson model in the localized regime.  To give full details of this argument would consume considerable
space, much of which would more or less reproduce earlier published work with only a few minor changes.
For this reason, we will be less detailed than might otherwise be desirable.


We believe that under the assumptions set out above, $\psi_{n-1}(\tfrac xn)$ converges to $2\pi$
times a Poisson jump process.  Further, it appears that in the natural scaling limit with $x$
fixed, the process $\psi_{k}(\tfrac xn)$ also converges to ($2\pi$ times) a Poisson jump process
(cf. \cite{KillipNakano}).  It would be nice to deduce these facts directly from the recurrence equation;
we have not done this, but hope that someone will pursue these matters.  This approach to the problem would be
closer to the original approach to the one dimensional Anderson model, \cite{Molchanov}, although estimates
on the size of the eigenfunctions are used even there.

The basis of Minami's work is a limit law for a Poisson process---the analogue of the convergence
of the binomial distribution to a Poisson distribution.   Specifically, suppose that for each $n$ we have
independent point processes $d\eta_{j,n}$ for $1\leq j \leq n$ and let us form the union,
$d\xi_n:=\sum_j d\eta_{j,n}$.  Then $d\xi_n$ converges to the Poisson process with intensity $d\nu$ provided\\
\hspace*{\parindent}(a) $d\eta_{j,n}$ is a null array, that is,
    $\sup_j \Prob\{\eta_{j,n}(I) > 0\} \to 0$ as $n\to\infty$,\\
\hspace*{\parindent}(b) $\sum_j \Prob\{\eta_{j,n}(I) \geq 2 \} \to 0$ as $n\to\infty$, and\\
\hspace*{\parindent}(c) $\sum_j \Prob\{\eta_{j,n}(I) =1 \} \to \nu(I)$ as $n\to\infty$,\\
for each compact set $I$.  This result is Corollary~7.5 in \cite{Kallenberg} and can be found in most books
on point processes.

This limit theorem appears here in the following manner.  Let us break the $n\times n$ CMV matrix $\C^{(n)}$
into $\sim\log[n]$ blocks of equal length $N\sim n/\log[n]$ by introducing independent random unitary boundary
conditions (just as we did in the Introduction in order to obtain an $n\times n$ matrix from the originally
semi-infinite one).  In this way, we obtain the direct sum of statistically independent matrices, so the eigenvalue
process is the union of $\sim n/\log[n]$ independent point processes.  This leaves us with two jobs: to check the
hypotheses of the theorem above (this is quick) and to show that the introduction of these boundary conditions has not
significantly affected the eigenvalue statistics (this is anything but quick).

By choosing the boundary conditions uniformly from the circle, the distribution of the eigenvalues of each
$N\times N$ block is rotationally invariant.  This resolves the hypothesis~(a) above and will suffice
to justify~(c) once we prove~(b).  To verify~(b) we invoke the following result from Section~6.1
of \cite{StoiciuPhD}.

\begin{lemma}\label{L:two evs}  Given any CMV matrix of size $N\times N$ with Verblunsky coefficients that are
independent and follow rotationally invariant laws, one has
\begin{equation}
\Prob(\Omega) \leq \tfrac12 L^2 N^2,
\end{equation}
where $\Omega$ is the event that two or more eigenvalues belong to a fixed arc of length $2\pi L$.
(Note that the expected number of eigenvalues in this arc is $LN$.)
\end{lemma}

This settles the first of our two jobs.  The second is to explain why the introduction of boundary conditions
does not completely rearrange the eigenvalues.  The reason is simple: the eigenfunctions are very strongly
localized with sharply decaying tails.  Proving this fact is not so simple; however, there is a wealth of
work on the Anderson model (i.e.,  for related operators with non-decaying random potentials)
that provides guidance.  We will be studying fractional moments of the Green's function,
which is an idea of \cite{AM}; while our approach to estimating the transfer matrix is remeniscent
of \cite{KunzSouillard,ShubVakWolff}.

For most of the remainder of this section, that is, until Corollary~\ref{C:finite vol},
we will be dealing with the full semi-infinite CMV matrix $\C$.

The Szeg\H{o} recurrence for the orthonormal polynomials can be encoded with a matrix:
$$
A_k=\frac{1}{\sqrt{1-|\alpha_k|^2}} \begin{bmatrix} z & -\bar\alpha_k \\ -\alpha_kz & 1 \end{bmatrix}.
$$
For $l \geq k$, we define the transfer matrices
$$
T(l,k) = A_{l-1}A_{l-2}\cdots A_{k+1} A_k,
$$
with $T(k,k)=\Id$.  Notice that both $A_k$ and $T(l,k)$ depend on $z\in\Cmplx$; however we suppress
this in the notation, since invariably $z$ will be fixed.

\begin{lemma}\label{A lemma}
Let $\alpha_k$ follow a rotationally invariant law, then
\begin{equation}
\Exp\bigl\{ \|A_k v \|^{-1} \bigr\} \leq \|v\|^{-1} \Exp\bigl\{1-\tfrac14|\alpha_k|^2\bigr\}
\end{equation}
for any non-zero vector $v\in \Cmplx^2$ and any $|z|=1$.
\end{lemma}

\begin{proof}
Clearly it suffices to prove this result when $v=(x,y)$ is a unit vector.  In this case, a little
computation shows
$$
\|A_k v \|^{2} = \frac{1-4\Re(x\bar{y}z\alpha_k)+|\alpha_k|^2}{1-|\alpha_k|^2}.
$$
The lemma now degenerates to the estimation of an integral.  We will use following notation:
$r=|\alpha_k|$, $R=2|x\bar{y}|$, $\theta=\arg(x\bar{y}\alpha_k)$.  Note that $R,r\in[0,1]$, in the former
case, because $|x|^2+|y|^2=1$.
\begin{align}
\int_0^{2\pi} \frac{\sqrt{1-r^2}}{\sqrt{1-2Rr\cos(\theta)+r^2}}\,\frac{d\theta}{2\pi}
&\leq \frac2\pi \int_0^{2\pi} \frac{\sqrt{1-r^2}}{\sqrt{1-2r\cos(\theta)+r^2}}\,\frac{d\theta}{2\pi} \\
&= \frac2\pi \frac{\sqrt{1-r^2}}{1+r} \, \mathbf{K}\!\left(\tfrac{2\sqrt{r}}{1+r} \right) \leq 1 -\tfrac14 r^2.
\end{align}
where $\mathbf{K}(k)$ is the complete elliptic integral of modulus $k$.  The first inequality
here follows from the fact that this integral is an increasing function of $R$; to see this
differentiate and compare the resulting integrand at $\theta$ and $\theta+\pi$.  To obtain the
second inequality with the factor $\frac14$ takes some trouble; however,
the fact that it is true with some small coefficient follows easily by studying the behaviour as $r\downarrow 0$.
This simpler fact is actually sufficient for our purposes.
\end{proof}

By applying this inductively, we obtain results for the transfer matrices:

\begin{lemma}\label{T lemma}
Suppose the Verblunsky coefficients are independent and follow rotationally invariant laws.
Given $0\leq s \leq 1$ and $|z|=1$,
\begin{equation}\label{TLemma1}
\Exp\Bigl\{ \| T(l,k) \|^{-s} \Bigr\} \leq \exp\biggl[ -\tfrac{s}4 \sum_{j=k}^{l-1}
    \Exp\bigl\{ |\alpha_j|^2 \bigr\} \biggr]
\end{equation}
for any $k\leq l$.  More generally,
\begin{equation}\label{TLemma2}
\Exp\Bigl\{ \tfrac{\| T(l,k) \|^s}{\| T(r,k) \|^s} \Bigr\}
    \leq \exp\biggl[ -\tfrac{s}4 \sum_{j=k}^{r-1} \Exp\bigl\{ |\alpha_j|^2 \bigr\} \biggr]
\end{equation}
for $k\leq l\leq r$.
\end{lemma}

\begin{proof}
Let us first remark that \eqref{TLemma1} is the special case of \eqref{TLemma2} when $k=l$.  We will prove a
slightly more general result:
\begin{equation}\label{TLemma3}
\Exp\Bigl\{ \| T(r,l)v \|^{-s}\|v\|^s \Bigr\}
        \leq \exp\biggl[ -\tfrac{s}4 \sum_{j=l}^{r-1} \Exp\{ |\alpha_j|^2 \} \biggr]
\end{equation}
where $v$ is a random unit vector that is statistically independent of $\alpha_j$ for $l\leq j \leq r-1$.  Note that
the result for general $s$ will follow from $s=1$ by Jensen's inequality.

Clearly \eqref{TLemma3} is well formulated for proof by induction. When $r=l$ one has equality.  Assuming the
result for $T(r,l)$, we have
\begin{align}
\Exp\Bigl\{ \| T(r+1,l)v \|^{-1}\|v\|^s \Bigr\} &=  \Exp\Bigl\{ \| A_r T(r,l)v \|^{-1}\|v\|^s \Bigr\} \\
&\leq \Exp\bigl\{1-\tfrac14|\alpha_k|^2\bigr\} \Exp\Bigl\{ \| T(r,l)v \|^{-1}\|v\|^s \Bigr\}
\end{align}
by Lemma~\ref{A lemma}.  To finish the proof, notice that $1-\frac{x}4 \leq e^{-x/4}$ for $x\in[0,\infty)$.
\end{proof}

\begin{prop}\label{P:Osc}
Fix $0< s\leq\min\{\frac12s_0,\frac14 \}$.  With probability one, there exists a (random) unit vector $v$ so
that $T(l,k)v\to 0$ as $l\to\infty$; moreover, there exists $c>0$ so that
\begin{align}
\Exp\bigl\{ \|T(l,k)v\|^s \bigr\} \lesssim (k+1) \exp\!\big[ -c(l^\epsilon-k^\epsilon) \bigr]
\end{align}
uniformly for $0\leq k \leq l$ and $|z|=1$.
\end{prop}

\begin{proof}
We simply take expectations in the proof of the Osceledec Theorem. This theorem is covered in
many textbooks, for example, \cite[\S10.5]{SimonOPUC2}, which can be consulted for more detail on what
follows.

Let $P_l$ denote the orthogonal projection onto the contracting subspace of $T(l,k)\in SL(2,\Cmplx)$; that is,
the subspace where $\|Tu\|=\|T\|^{-1}\|u\|$.  (This is one dimensional when $\|T\|>1$.)  Then
\begin{equation}\label{P diff}
\|P_{l+1}-P_l\|\leq \frac{\|A_{l}\|^2}{\|T(l,k)\|^2}.
\end{equation}
This is a ratio of independent random variables. As $\|A_l\|^2\leq 2(1-|\alpha_l|^2)^{-1}$, it follows from
\eqref{IIIass1} that $\Exp\{ \|A_{l}\|^{2s} \} \lesssim 1$.  (This implicit constant depends on $s$ as
do many more below.) Therefore, by applying Lemma~\ref{T lemma} and \eqref{IIIass2} we see that there
exists $c>0$ so that
$$
\Exp\bigl\{  \|P_{l+1}-P_l\|^{s} \bigr\} \lesssim \exp\biggl[ -\tfrac{s}2 \sum_{j=k}^{l-1} \Exp\{ |\alpha_j|^2 \} \biggr]
\lesssim \exp\!\big[ -c(l^\epsilon-k^\epsilon) \bigr]
$$
for any $s\leq\min\{s_0,\frac12 \}$.  This implies that $P_{l+1}-P_l$ is almost surely summable and so
$P_\infty:=\lim_{l\to\infty} P_l$ exists.

Let $v$ be a unit vector from the range of $P_\infty$.  We must now study $\|T(l,k)v\|$.
Using only elementary facts, we have
\begin{align}
\Exp\{ \|T(l,k)v\|^s \} &= \Exp\{ \|T(l,k)P_\infty\|^s \} \\
&\leq \Exp\{ \|T(l,k)P_l\|^s \} + \sum_{r=l}^\infty \Exp\{ \|T(l,k)\|^s \|P_{r+1}-P_r\|^{s} \} \\
&\leq \Exp\{ \|T(l,k)\|^{-s} \} + \sum_{r=l}^\infty \Exp\{ \|T(l,k)\|^s \|P_{r+1}-P_r\|^{s} \}.
\end{align}
By Lemma~\ref{T lemma} and \eqref{IIIass2},
$$
\Exp\bigl\{ \|T(l,k)\|^{-s} \bigr\} \lesssim \exp\biggl[
    -\tfrac{s}4 \sum_{j=k}^{l-1} \Exp\bigl\{ |\alpha_j|^2 \bigr\} \biggr]
\lesssim \exp\!\big[ -c(l^\epsilon-k^\epsilon) \bigr]
$$
for some $c>0$. Next we deal with the sum: using \eqref{P diff}, the Cauchy--Schwarz inequality,
and the independence of the Verblunsky coefficients, we have
\begin{align}
\Exp\bigl\{ \|T(l,k)\|^s &\|P_{r+1}-P_r\|^{s} \bigr\}^2 \\
\leq{} & \Exp\{ \|A_r\|^{4s} \}\Exp\{ \|T(r,k)\|^{-2s} \}
\Exp\biggl\{ \frac{\|T(l,k)\|^{2s} }{ \|T(r,k)\|^{2s} } \biggr\}.
\end{align}
The first factor is uniformly bounded by \eqref{IIIass1}.  The second and third factors can be estimated
via Lemma~\ref{T lemma}.  In this way, we obtain
$$
\Exp\bigl\{ \|T(l,k)\|^s \|P_{r+1}-P_r\|^{s} \bigr\}
\lesssim \exp\!\big[ -c(2r^\epsilon - l^\epsilon-k^\epsilon) \bigr]
$$
for some $c>0$.  This is clearly summable in $r$; indeed
$$
\sum_{r=l}^\infty \exp\bigl[-2cr^\epsilon\bigr] \lesssim  (l+1)^{1-\epsilon} \exp\bigl[ -2cl^\epsilon \bigr].
$$
Consequently,
$$
\sum_{r=l}^\infty \Exp\bigl\{ \|T(l,k)\|^s \|P_{r+1}-P_r\|^{s} \bigr\}
\lesssim (l+1) \exp\!\big[ -cs(l^\epsilon-k^\epsilon) \bigr].
$$

Collecting all the estimates we have derived, we obtain the existence of a $c>0$ so that
$$
\Exp\bigl\{ \|T(l,k)v\|^s \bigr\} \lesssim (l+1) \exp\!\big[ -c(l^\epsilon-k^\epsilon) \bigr].
$$
By reducing $c$ a little bit, one may replace the polynomial factor by $(k+1)$ as written in the statement.
\end{proof}

It is presumably possible to obtain much more precise estimates for the decay of $\|T(l,k)v\|$ using the
methods from \cite{KLS} (cf. \cite[\S 12.7]{SimonOPUC2}).  While their techniques are strong, they
pursue only the almost sure behaviour, which is why we dragged the reader through the mud here.

In order to proceed, we need to upgrade Proposition~\ref{P:Osc} to an estimate on the Green's function
(i.e., the matrix elements of the resolvent),
$$
G_{kl}(z) := (\C-z)^{-1}_{kl},
$$
for $|z|=1$.  From \cite[\S4.4]{SimonOPUC1}, we learn the following:
\begin{equation}
G_{kl}(z) = \begin{cases}
(2z)^{-1} \chi_l(z)p_k(z) & k>l \text{ or } k=l \text{ is odd} \\
(2z)^{-1} x_k(z) \pi_l(z) & l>k \text{ or } k=l \text{ is even}.
\end{cases}
\end{equation}
Here
\begin{gather*}
\begin{bmatrix}p_{2l}\\\pi_{2l}\end{bmatrix} = z^{-l} T(2l+1,0) \begin{bmatrix} F(z) + 1 \\ F(z) - 1  \end{bmatrix},\ \
\begin{bmatrix}z^{-1} \pi_{2l-1} \\p_{2l-1}\end{bmatrix}
        = z^{-l} T(2l,0) \begin{bmatrix} F(z) + 1 \\ F(z) - 1  \end{bmatrix},\\[3mm]
\begin{bmatrix}x_{2l}\\\chi_{2l}\end{bmatrix} = z^{-l} T(2l+1,0) \begin{bmatrix} 1 \\ 1  \end{bmatrix},
\text{ and }
\begin{bmatrix}z^{-1}\chi_{2l-1}\\x_{2l-1}\end{bmatrix} = z^{-l} T(2l,0) \begin{bmatrix} 1 \\ 1  \end{bmatrix},
\end{gather*}
where $F(z)$ is the Carath\'eodory function given in \eqref{E:Carat}.  Note that $x_k(z)$ is the complex conjugate of
$\chi_k(1/\bar{z})$; in particular, these two sequences have the same moduli when $z$ lies on the unit circle.

For $|z|=1$, $F(z)$ is purely imaginary with probability one because $\C$ has purely point spectrum
with probability one.  (For the latter fact, see \cite[Theorem~12.7.5]{SimonOPUC2} or combine the Simon--Wolff criterion,
\cite[Theorem 10.2.1]{SimonOPUC2}, with Proposition~\ref{P:Osc}.)  This implies that $p_k(z)$ and $\pi_k(z)$ are complex
conjugates for such $z$.

The way we have introduced $p_k$ and $\pi_k$ belies their true significance; they are the Weyl
solutions---they are proportional to the decaying solutions (whenever such solutions exist).  In this way we find that
for $|z|=1$ and $k\leq l$,
\begin{equation}\label{E:ratios}
\left|\frac{p_l(z)}{p_k(z)}\right| = \left|\frac{\pi_l(z)}{\pi_k(z)}\right| = \| T(l,k) v\|
\end{equation}
with probability one, provided $v$ is chosen as in Proposition~\ref{P:Osc}.

\begin{coro}
Fix $0< s < 1$.  There exists $c>0$ so that
\begin{align}\label{EC:decay}
\Exp\bigl\{ \bigl|G_{kl}(z)\bigr|^s + \bigl|G_{lk}(z)\bigr|^s\bigr\}
        \lesssim (k+1) \exp\!\big[ -c(l^\epsilon-k^\epsilon) \bigr]
\end{align}
uniformly for $0\leq k \leq l$ and $|z|=1$.  Moreover, the same estimate holds uniformly in
$|z|\leq 1$ for a reduced  value of $c$.
\end{coro}

\begin{proof}
By employing Kolmogorov's theorem, \cite{Duren}, as in \cite[Lemma~4.1.1]{StoiciuPhD}, we see that
\begin{equation}\label{E:Kolm}
\Exp\bigl\{ \bigl|G_{kl}(z)\bigr|^s \bigr\} \leq 2^{2-s} \sec(s\pi/2).
\end{equation}
for any choice of $k,l$ and $|z|\leq 1$.  This shows us that it is sufficient to prove the result for $s$ small,
since the general version can then be recovered by interpolation (or rather, by H\"older's inequality).

Let us begin with $|z|=1$. From the discussion preceding this corollary, we have
\begin{equation}\label{E:ratios2}
\left|\frac{G_{kl}(z)}{G_{kk}(z)}\right| = \left|\frac{G_{lk}(z)}{G_{kk}(z)}\right| = \| T(l,k) v\|
\end{equation}
and so by the Cauchy--Schwartz inequality,
\begin{equation}\label{E:ratios3}
\Exp\bigl\{ \bigl|G_{kl}(z)\bigr|^s + \bigl|G_{lk}(z)\bigr|^s\bigr\}
\leq 2 \Exp\bigl\{ \| T(l,k) v\|^{2s} \bigr\} \Exp\bigl\{ |G_{kk}(z)|^{2s} \bigr\}.
\end{equation}
The first factor here was estimated in Proposition~\ref{P:Osc}, while the second can be bounded
using \eqref{E:Kolm}.  This proves \eqref{EC:decay}.

To extend the result to $z\in\Disk$, one may apply the argument from
Theorem~4.2 in \cite{ASFH}.
\end{proof}

\begin{coro}\label{C:finite vol}
Fix $0< s < 1$.  There exists $c>0$ and $A>0$ so that
\begin{align}
\Exp\bigl\{ \bigl| (\C^{(n)}-z)^{-1}_{kl} \bigr|^s \bigr\}
        \leq  A \, n \exp\!\big[ -cn^{\epsilon-1}|l-k| \bigr]
\end{align}
for all $|z|\leq 1$ and $n$.
\end{coro}

\begin{proof}
This result for finite matrices follows from the previous corollary by applying the resolvent
identity as in Lemma~4.2.3 of \cite{StoiciuPhD}.  To make the final result look cleaner, we
employed the estimate
$$
|l^\epsilon-k^\epsilon| \geq \epsilon n^{\epsilon-1} |k-l|, \quad\forall\ k,l\in[0,n],
$$
which follows from the lower bound on the derivative of $x\mapsto x^\epsilon$.
\end{proof}

With Lemma~\ref{L:two evs} and Corollary~\ref{C:finite vol} as input, one may now simply repeat
the analysis of \cite{Minami} to prove part (iii) of Theorem~\ref{T:main}.  This input is
also sufficient to prove Poisson statistics for the locations of the eigenfunctions
as in \cite{KillipNakano}.


\end{document}